\documentclass{article}
\usepackage{amsfonts}

\usepackage{graphicx}
\usepackage{amsmath}

\newenvironment{proof}[1][Proof]{\textbf{#1.} }{\ \rule{0.5em}{0.5em}}
\input{tcilatex}

\begin{document}

\title{Quantum Stratonovich Calculus and the Quantum Wong-Zakai Theorem}
\author{John Gough \\
School of Computing and Informatics,\\
Nottingham Trent University,\\
Nottingham, NG1 4BU, United Kingdom}
\date{}
\maketitle

\begin{abstract}
We extend the It\={o}-to-Stratonovich analysis or quantum stochastic
differential equations, introduced by Gardiner and Collett for emission
(creation), absorption (annihilation) processes, to include scattering
(conservation) processes. Working within the framework of quantum stochastic
calculus, we define Stratonovich calculus as an algebraic modification of
the It\={o} one and give conditions for the existence of Stratonovich
time-ordered exponentials. We show that conversion formula for the
coefficients has a striking resemblance to Green's function formulae from
standard perturbation theory. We show that the calculus conveniently
describes the Markov limit of regular open quantum dynamical systemsin much
the same way as in the Wong-Zakai approximation theorems of classical
stochastic analysis. We extend previous limit results to multiple-dimensions
with a proof that makes use of diagrammatic conventions.
\end{abstract}

\bigskip

\section{Introduction}

Quantum stochastic calculus \cite{BSW}-\cite{GardColl} was developed as a
framework to construct concrete models of irreversible quantum dynamical
systems. Prior to this, models tended to consider a system couple to an
environment, but with only the system being accessible to physical
measurement: as a result the environment observables were often relegated to
a secondary status, leaving one with a master equation for the state of the
system only \cite{Davies}, \cite{Spohn}.

Hudson and Parthasarathy \cite{HP} in 1984 presented a rigorous theory of
integration with respect to processes on Bosonic (later Fermionic) Fock
spaces generalizing the It\={o}-Doob theory of stochastic integration. In
addition to integrals with respect to time, they also introduced integrals
with respect to creation, annihilation and number (more generally,
scattering) processes. Motivated by non-commutative Feynman-Kac formulae,
they were able to describe unitary dynamical evolutions of a system coupled
to the Fock space environment which reduced to an irreversible Markov
dynamics for the system when averaged over the partial trace with respect to
the Fock vacuum. Here the Schr\"{o}dinger equation is replaced by a quantum
stochastic differential equation (QSDE) driven by the creation, annihilation
and scattering processes.

Independently, Gardiner and Collett \cite{GardColl} in 1985 gave the version
of quantum stochastic integration for the Bosonic creation and annihilation
that is best known amongst the physics community. Although they did not
include the scattering processes, they did introduce several important
physical concepts, in particular, they gave to the noise the status of a
physical observable. This has been vital in subsequent analysis of quantum
filtering and feedback where the environment can act as an apparatus/
communication channel \cite{Milburn Wiseman}-\cite{bouten et al}. In their
analysis, they also introduced the Stratonovich version of the theory by
extending the usual mid-point definition to non-commuting processes. This is
a natural physical choice for two reasons: unlike the It\={o} form, the
Leibniz rule of differential calculus holds for Stratonovich differentials
and so physical symmetries are more apparent; secondly, in classical
analysis it is generally the case that if the model can be obtained as a
singular limit of regular dynamical models, then it is the Stratonovich form
that resembles the pre-limit equations the most. Historically, it was
actually the latter reason that lead to Stratonovich initially introducing
his modification of the It\={o} theory. In ordinary stochastic analysis,
such results involving a central limit effect for stochastic processes are
known as Wong-Zakai theorems. Our original motivation stems from quantum
Markov limits \cite{AFL} - \cite{GQCLT} and the desire to understand the
limit processes in a Stratonovich sense.

To give a concrete mathematical account, we start from the quantum It\={o}
theory developed by Hudson and Parthasarathy, and deduce quantum
Stratonovich calculus as an algebraic modification of the quantum It\={o}
one which restores the Leibniz rule. An approach starting from Gardiner and
Collett's input processes would have been more appealing from a physical
point of view, however, we do not want to bypass questions of the
mathematical status of the objects considered. Traditionally, the
Stratonovich integral is defined through a midpoint Riemann sum
approximation: this has been extended by Chebotarev \cite{Cheb} to quantum
stochastic integrals, however, we emphasize that our formulation here is to
define Stratonovich integrals as combinations of well-defined It\={o}
integrals where possible. Rather than there been an unique version, we find
that there are degrees of freedom in how we actually achieve this\ - we
refer to this as a ``gauge'' freedom - and that the standard (symmetric)
choice, corresponding to the midpoint rule, is just one possibility. We give
the self-consistency formula $\mathbf{G}=\mathbf{G}_{0}+\mathbf{G}_{0}%
\mathbf{VG}$ relating the matrix of It\={o} coefficients $\mathbf{G}$\textbf{%
\ }to the matrix of Stratonovich coefficients $\mathbf{G}_{0}$. (Here the
``potential'' $\mathbf{V}$ is half the noise covariance matrix plus the
gauge.) Rather surprisingly, this has the same algebraic form as the one
relating the free and perturbed Green's functions in scattering theory: a
fact that we readily exploit. It is shown that the It\={o} coefficients of a
unitary process are related to Hamiltonian Stratonovich coefficients $\left( 
\mathbf{G}_{0}^{\dag }=-\mathbf{G}_{0}\right) $ and that this is true for
any gauge so long as the self-consistency formula can be solved. This allows
the interpretation that the Stratonovich calculus can be viewed as a
perturbation of the It\={o} calculus, and \textit{vice versa}.

A major motivation here is the results of \cite{GQCLT} for a quantum Markov
limit involving emission, absorption and scattering. We formulate the limit
as a Wong-Zakai result where the Stratonovich QSDE resembles the pre-limit
Schr\"{o}dinger equation (with gauge set by the imaginary part of the
complex damping). Our key requirement is convergence of the ``Neumann
series'' $\mathbf{G}=\sum_{n=0}^{n}\mathbf{G}_{0}\left( \mathbf{VG}%
_{0}\right) ^{n}$. We generalize the result to multiple channel noise
sources and make the proof more accessible by means of diagrammatic
conventions which make the connections with the Dyson series expansion
transparent. We also present the results in a fluxion notation, as an
alternative to the differential increment language, which is closer to the
formulation employed by Gardiner and Collett \cite{GardColl} and effectively
generalizes their results to include scattering. This also reveals a new
representation (lemma 3) for the Evans-Hudson flow maps.

\section{Quantum Stochastic Calculus}

\subsection{Quantum Processes}

Given a Hilbert space $\frak{h}_{1}$, the Fock space over $\frak{h}_{1}$ is
the Hilbert space $\Gamma \left( \frak{h}_{1}\right) $ spanned by
symmetrized $n$-particle vectors $\varphi _{1}\hat{\otimes}\cdots \hat{%
\otimes}\varphi _{n}:=\left( n!\right) ^{-1}\sum_{\sigma \in \frak{S}%
_{n}}\varphi _{\sigma \left( 1\right) }\otimes \cdots \otimes \varphi
_{\sigma \left( n\right) }$ for $n\geq 0$ arbitrary, $\varphi _{j}\in \frak{h%
}_{1}$ and $\frak{S}_{n}$ the group of permutations on $n$ labels. The
special case $n=0$ requires the introduction of unit vector $\Omega $ called
the Fock vacuum vector. The inner product on $\Gamma \left( \frak{h}%
_{1}\right) $ is given by $\left\langle \varphi _{1}\hat{\otimes}\cdots \hat{%
\otimes}\varphi _{n}|\psi _{1}\hat{\otimes}\cdots \hat{\otimes}\psi
_{n}\right\rangle =\delta _{nm}\left( n!\right) ^{-1}\sum_{\sigma \in \frak{S%
}_{n}}\left\langle \varphi _{1}|\psi _{\sigma (1)}\right\rangle \cdots
\left\langle \varphi _{n}|\psi _{\sigma (n)}\right\rangle $.

The exponential vector with test function $\varphi \in \frak{h}_{1}$ is
defined to the Fock space vector 
\begin{equation}
\varepsilon \left( \varphi \right) =\sum_{n=0}^{\infty }\frac{1}{\sqrt{n!}}%
\underset{n\text{ fold}}{\underbrace{\varphi \otimes \cdots \otimes \varphi }%
},
\end{equation}
and we have $\left\langle \varepsilon \left( \varphi \right) |\varepsilon
\left( \psi \right) \right\rangle =\exp \left\langle \varphi |\psi
\right\rangle $. If $S$ is a dense subset of $\frak{h}_{1}$ then the vectors 
$\varepsilon \left( \varphi \right) $, with $\varphi \in S$, are total in $%
\Gamma \left( \frak{h}_{1}\right) $, that is, the closure of the span of
these vectors gives the whole Fock space.

We now take the one-particle space to be the Hilbert space of $\mathbb{C}%
^{N} $-valued square-integrable functions of positive time: this consists of
measurable functions $\mathbf{f}=\left( f_{1},\cdots ,f_{n}\right) $ with $%
\int_{0}^{\infty }\sum_{i=1}^{n}|f_{i}\left( t\right) |^{2}<\infty $. An
orthogonal projection $\Pi _{s}$ is defined on the one-particle space for
each $s>0$ by taking $\Pi _{s}\mathbf{f}=\left( \chi _{\left[ 0,s\right]
}f_{1},\cdots ,\chi _{\left[ 0,s\right] }f_{n}\right) $ where $\chi _{\left[
0,s\right] }$ is the indicator function for the interval $\left[ 0,s\right] $%
. An $N$-channel quantum noise source is modelled by operators processes $%
\left\{ A_{t}^{\alpha \beta }:t>0\right\} $ acting on the corresponding Fock
space $\frak{F}$. For $\alpha ,\beta \in \left\{ 0,1,\cdots ,n\right\} $,
these processes are defined on the domain of exponential vectors by 
\begin{equation}
\left\langle \varepsilon \left( \mathbf{f}\right) \left| \left\{
A_{t}^{\alpha \beta }-\int_{0}^{t}f_{\alpha }^{\ast }\left( s\right)
g_{\beta }\left( s\right) ds\right\} \right. \varepsilon \left( \mathbf{g}%
\right) \right\rangle =0
\end{equation}
where we include the index zero by setting $f_{0}=g_{0}=1$. We shall adopt
the convention that lower case Latin indices (with the exception of $t$ and $%
s$ which we reserve for time!) range over the values $1,\cdots ,N$ while
lower case Greek indices range over $0,1,\cdots ,N$. We also apply an
Einstein summation convention for repeated indices over the appropriate
range.

We also fix a Hilbert space $\frak{h}$, called the initial space. Let $D$ be
a domain in $\frak{h}$ and take $S$ to be the space of bounded $\mathbb{C}%
^{N}$-valued functions. A family $X_{\cdot }=\left\{ X_{t}:t\geq 0\right\} $
of operators on $\frak{h}\otimes \frak{F}$ is said to be an adapted quantum
stochastic process based on $\left( D,S\right) $ if, for each $t\geq 0$, $%
X_{t}u\otimes \varepsilon \left( \mathbf{f}\right) $ is defined for each $%
u\in D$ and $\mathbf{f}=\left( f_{1},\cdots ,f_{n}\right) \in S$ and is
independent of the values $f_{i}\left( s\right) $ for $s>t$. If $X_{\cdot
}^{\alpha \beta }$ are adapted processes, their stochastic integral $%
X_{\cdot }$\ may be written as (implied summation!) $X_{t}=\int_{0}^{t}a_{%
\alpha }^{\dag }\left( s\right) X_{s}^{\alpha \beta }a_{\beta }\left(
s\right) ds$, with the meaning that 
\begin{multline}
\left\langle u\otimes \varepsilon \left( \mathbf{f}\right) \left|
\int_{0}^{t}a_{\alpha }^{\dag }\left( s\right) X_{s}^{\alpha \beta }a_{\beta
}\left( s\right) ds\right. v\otimes \varepsilon \left( \mathbf{g}\right)
\right\rangle  \notag \\
:=\int_{0}^{t}f_{\alpha }^{\ast }\left( s\right) \left\langle u\otimes
\varepsilon \left( \mathbf{f}\right) \left| X_{s}^{\alpha \beta }v\otimes
\varepsilon \left( \mathbf{g}\right) \right. \right\rangle g_{\beta }\left(
s\right) ds
\end{multline}
The stochastic integral is more often written in the form $%
X_{t}=\int_{0}^{t}X_{s}^{\alpha \beta }dA_{s}^{\alpha \beta }$ or
equivalently $\int_{0}^{t}dA_{s}^{\alpha \beta }\,X_{s}^{\alpha \beta }$
with these integrals making sense as Riemann-It\={o} limits for locally
square-integrable integrands. The stochastic integral will again be an
adapted process. At the moment, the symbols $a_{\alpha }^{\dag }\left(
t\right) ,a_{\beta }\left( t\right) $ have no meaning other than notational,
however $a_{0}^{\dag }\left( t\right) $ and $a_{0}\left( t\right) $ are
easily interpreted as the identity operator and $a_{i}\left( t\right) $ as a
pointwise Malliavin gradient. What is crucial is that they appear in Wick
order: that is, $a_{\alpha }^{\dag }\left( t\right) $ to the left of $%
a_{\beta }\left( t\right) $. Let $Y_{t}=\int_{0}^{t}a_{\alpha }^{\dag
}\left( s\right) Y_{s}^{\alpha \beta }a_{\beta }\left( s\right) ds$ be a
second integral, with the $Y_{\cdot }^{\alpha \beta }$ adapted, then we have
the formula \cite{HP} 
\begin{eqnarray*}
X_{t}Y_{t} &=&\int_{0}^{t}a_{\alpha }^{\dag }\left( s\right) X_{s}^{\alpha
\beta }Y_{s}a_{\beta }\left( s\right) ds+\int_{0}^{t}a_{\alpha }^{\dag
}\left( s\right) X_{s}Y_{s}^{\alpha \beta }a_{\beta }\left( s\right) ds \\
&&+\int_{0}^{t}a_{\alpha }^{\dag }\left( s\right) X_{s}^{\alpha \mu }P_{\mu
\nu }Y_{s}^{\nu \beta }a_{\beta }\left( s\right) ds,
\end{eqnarray*}
where we introduce 
\begin{equation}
P^{\mu \nu }=\left\{ 
\begin{array}{cc}
1, & \mu =\nu \neq 0; \\ 
0, & \text{otherwise.}
\end{array}
\right.
\end{equation}

Introducing the differential notation $dX_{t}=X_{t}^{\alpha \beta
}dA_{t}^{\alpha \beta }$, etc., we may write the quantum It\={o} formula in
the more familiar guise as 
\begin{equation}
d\left( X_{t}Y_{t}\right) =\left( dX_{t}\right) Y_{t}+X_{t}\left(
dY_{t}\right) +\left( dX_{t}\right) \left( dY_{t}\right)
\end{equation}
where the It\={o} correction is $\left( dX_{t}\right) \left( dY_{t}\right)
=X_{t}^{\alpha i}Y_{t}^{i\beta }dA_{t}^{\alpha \beta }\equiv X_{t}^{\alpha
\mu }P^{\mu \nu }Y_{t}^{\nu \beta }dA_{t}^{\alpha \beta }$. Let us define
iterated integrals in the natural way: 
\begin{multline}
\left\langle \varepsilon \left( \mathbf{f}\right) \right. |\,\int_{\Delta
_{n}\left( t\right) }dA_{t_{n}}^{\alpha _{n}\beta _{n}}\cdots
dA_{t_{1}}^{\alpha _{1}\beta _{1}}\left. \,\varepsilon \left( \mathbf{g}%
\right) \right\rangle \\
=\int_{\Delta _{n}\left( t\right) }f_{\alpha _{n}}^{\ast }\left(
t_{n}\right) \cdots f_{\alpha _{1}}^{\ast }\left( t_{1}\right) g_{\beta
_{n}}\left( t_{n}\right) \cdots g_{\beta _{1}}\left( t_{1}\right)
\,\left\langle \varepsilon \left( \mathbf{f}\right) |\,\varepsilon \left( 
\mathbf{g}\right) \right\rangle
\end{multline}
where $\Delta _{n}\left( t\right) $ is the simplex $t\geq t_{n}>\cdots \geq
t_{1}>0$.

\subsection{Quantum Markov Evolutions}

Let $\left\{ G_{\alpha \beta }\right\} $ be bounded operators on a fixed
initial space $\frak{h}$, then there exist a unique solution $U_{\cdot }$\
to the equation $U_{t}=1+\int_{0}^{t}G_{\alpha \beta }U_{s}dA_{s}^{\alpha
\beta }$ which we can naturally interpret as the QSDE $dU_{t}=G_{\alpha
\beta }U_{t}dA_{t}^{\alpha \beta }$, with $U_{0}=1$. In such cases, we may
write $U_{\cdot }$ as the \textit{Dyson-It\={o} time-ordered exponential} $%
U_{t}=\mathbf{\vec{T}}_{ID}\exp \int_{0}^{t}G_{\alpha \beta }dA^{\alpha
\beta }$.

\bigskip

\noindent \textbf{Proposition 1:} \emph{Let }$U_{\cdot }$\emph{\ be the
solution to the QSDE }$dU_{t}=G_{\alpha \beta }U_{t}dA_{t}^{\alpha \beta }$, 
\emph{with }$U_{0}=1$\emph{, where the }$\left\{ G_{\alpha \beta }\right\} $ 
\emph{are bounded operators on }$\frak{h}$\emph{. Necessary and sufficient
conditions \ \cite{HP}\ for }$U_{\cdot }$\emph{\ to be unitary process are
that }$G_{\alpha \beta }+G_{\beta \alpha }^{\dag }+G_{i\alpha }^{\dag
}G_{i\beta }=0=G_{\alpha \beta }+G_{\beta \alpha }^{\dag }+G_{i\alpha
}G_{i\beta }^{\dag }$\emph{\ with the general solution } 
\begin{eqnarray*}
G_{ij} &=&W_{ij}-\delta _{ij};\quad G_{i0}=L_{i}; \\
G_{0j} &=&-L_{k}^{\dag }W_{kj};\quad G_{00}=-\frac{1}{2}L_{k}^{\dag
}L_{k}-iH.
\end{eqnarray*}
\emph{\ where }$W_{ij},L_{i}$\emph{\ and }$H$\emph{\ are bounded operators
on the initial space with }$H$\emph{\ self-adjoint and }$W_{ij}^{\dag
}W_{jk}=\delta _{ik}=W_{ij}W_{jk}^{\dag }$\emph{.}

\bigskip

\noindent \textbf{Proposition 2:} \emph{Let }$U_{\cdot }$\emph{\ be the
unitary process described above The corresponding flow map is given by }$%
J_{t}\left( X\right) :=U_{t}^{\dag }\left( X\otimes 1\right) U_{t}$ \emph{%
for bounded }$X$\emph{\ on the initial space. We find that }$J_{t}\left(
X\right) $\emph{\ satisfies the QSDE} 
\begin{equation*}
dJ_{t}\left( X\right) =J_{t}\left( \mathcal{L}_{\alpha \beta }\left(
X\right) \right) dA_{t}^{\alpha \beta },
\end{equation*}
\emph{where the Evans-Hudson maps \cite{EVHUD} are given by }$\mathcal{L}%
_{\alpha \beta }\left( X\right) :=XG_{\alpha \beta }+G_{\beta \alpha }^{\dag
}X+G_{i\alpha }^{\dag }XG_{i\beta }$\emph{. }

\bigskip

Following Lindblad \cite{Lindblad}, the dissipation of a linear map $%
\mathcal{L}$\ on the algebra of bounded operators on $\frak{h}$\ is defined
to be the bilinear mapping $\frak{D}\mathcal{L}:\left( X,Y\right) \mapsto 
\mathcal{L}\left( XY\right) -\mathcal{L}\left( X\right) Y-X\mathcal{L}\left(
Y\right) $.

\bigskip

\noindent \textbf{Proposition 3:} \emph{The Evans-Hudson maps for a unitary
flow satisfy }$\mathcal{L}_{\alpha \beta }\left( X\right) ^{\dag }=\mathcal{L%
}_{\beta \alpha }\left( X^{\dag }\right) $\emph{\ and their dissipation is
described by the equation} 
\begin{equation*}
\frak{D}\mathcal{L}_{\alpha \beta }\left( X,Y\right) =\mathcal{L}_{\alpha
\mu }\left( X\right) P^{\mu \nu }\mathcal{L}_{\nu \beta }\left( Y\right) 
\emph{.}
\end{equation*}

\subsection{Approximations}

For each $\lambda >0$, we set $a_{0}^{\#}\left( t,\lambda \right) =0$ and
take $a_{i}^{\#}\left( t,\lambda \right) =A^{\#}\left( \varphi \left(
i,t,\lambda \right) \right) $ where $\varphi \left( i,t,\lambda \right) $ is
a $\mathbb{C}^{N}$-valued square-integrable function on $[0,\infty )$: we
also take $t\rightarrow \varphi \left( i,t,\lambda \right) $ to be strongly
differentiable. We assume that $\left\langle \varphi \left( i,t,\lambda
\right) |\varphi \left( j,s,\lambda \right) \right\rangle \equiv
C_{ij}\left( t-s,\lambda \right) $ where $C_{ij}\left( \tau ,\lambda \right)
=C_{ji}\left( -\tau \right) ^{\ast }$ is integrable in $\tau $, and that we
have the convergence $\lim_{\lambda \rightarrow 0}C_{ij}\left( \tau ,\lambda
\right) =\delta _{ij}\,\delta \left( \tau \right) $ in the sense of Schwartz
distributions. We set $\kappa _{ij}=\lim_{\lambda \rightarrow
0}\int_{0}^{\infty }C_{ij}\left( \tau ,\lambda \right) d\tau $ and note the
identity $\kappa _{ij}+\kappa _{ji}^{\ast }=\delta _{ij}$.

The $a_{i}^{\#}\left( t,\lambda \right) $ are approximations to quantum
white noises. We may introduce integrated processes $A_{t}^{\alpha \beta
}\left( \lambda \right) :=\int_{0}^{t}a_{\alpha }^{\dag }\left( s,\lambda
\right) a_{\beta }\left( s,\lambda \right) ds$ which serve as approximations
to the fundamental processes. The approximation will be termed symmetric if $%
\kappa _{ij}=\frac{1}{2}\delta _{ij}$, but this is only a special case.
Defining smeared exponential vectors by 
\begin{equation}
\varepsilon _{\lambda }\left( \mathbf{f}\right) =\exp \left\{
\int_{0}^{\infty }\sum_{i=1}^{N}f_{i}\left( t\right) a_{i}^{\dag }\left(
t\right) dt\right\} \,\Omega ,
\end{equation}
we see that the limit 
\begin{equation*}
\lim_{\lambda \rightarrow 0}\left\langle \varepsilon _{\lambda }\left( 
\mathbf{f}\right) \right. |\,\int_{\Delta _{n}\left( t\right) }a_{\alpha
_{n}}^{\ast }\left( t_{n},\lambda \right) \cdots a_{\alpha _{1}}^{\ast
}\left( t_{1},\lambda \right) a_{\beta _{n}}\left( t_{n},\lambda \right)
\cdots a_{\beta _{1}}\left( t_{1},\lambda \right) \,\left. \varepsilon
_{\lambda }\left( \mathbf{g}\right) \right\rangle
\end{equation*}
coincides with $\left( 5\right) $.Therefore, whenever we consider the limit
of Wick ordered expressions, it doesn't matter whether we have the symmetric
approximation or not.

A term such as $\int_{\Delta _{2}\left( t\right) }a_{i}\left( t_{2},\lambda
\right) a_{j}^{\dag }\left( t_{1},\lambda \right) $ must, however, be put to
Wick order as $\int_{\Delta _{2}\left( t\right) }a_{j}^{\dag }\left(
t_{1},\lambda \right) a_{i}\left( t_{2},\lambda \right) $\ plus an
additional term $\int_{\Delta _{2}\left( t\right) }C_{ij}\left(
t_{2}-t_{2},\lambda \right) $ which converges to $\kappa _{ij}t$. As a rule,
expressions out of Wick order will have a limit that depends on the
constants $\kappa _{ij}$.

\subsection{Notation and Conventions}

Let $\mathbf{X}=\left\{ X_{\alpha \beta }\right\} $ be an $\left( N+1\right)
\times \left( N+1\right) $ matrix of bounded operators on some Hilbert
space. (Previously, we had the It\={o} coefficients which were operators on $%
\frak{h}\otimes \frak{F}$.) We shall adopt the matrix representation 
\begin{equation}
\mathbf{X}=\left( 
\begin{tabular}{l|l}
$X_{00}$ & $X_{01},X_{02},\cdots $ \\ \hline
$X_{10}$ &  \\ 
$X_{20}$ & \multicolumn{1}{|c}{$\mathsf{X}$} \\ 
$\vdots $ & 
\end{tabular}
\right)
\end{equation}
where $\mathsf{X}$ is the $N\times N$ sub-matrix consisting of the entries $%
\left\{ X_{ij}\right\} $.

For instance, $\mathbf{P}=\left\{ P^{\alpha \beta }\right\} $ will be a
projection operator: 
\begin{equation}
\mathbf{P}=\left( 
\begin{tabular}{l|l}
$0$ & $0$ \\ \hline
$0$ & $\mathsf{1}$%
\end{tabular}
\right) ,\;\mathbf{Q}=\mathbf{1-P}=\left( 
\begin{tabular}{l|l}
$1$ & $0$ \\ \hline
$0$ & $\mathsf{0}$%
\end{tabular}
\right) .
\end{equation}

The quantum It\={o} correction\ is therefore described by the matrix $%
\mathbf{X}\left( t\right) \mathbf{PY}\left( t\right) \equiv \left\{
X_{\alpha \mu }\left( t\right) P^{\mu \nu }Y_{\nu \beta }\left( t\right)
\right\} $\textbf{.}

Let $\left\{ G_{\alpha \beta }\right\} \equiv \mathbf{G}$ be a matrix with
bounded operators on $\frak{h}$ as entries, then the Dyson-It\={o}
time-ordered exponential $U_{t}=\mathbf{\vec{T}}_{DI}\left\{ \exp
\int_{0}^{t}G_{\alpha \beta }dA^{\alpha \beta }\right\} $ will be unitarity
if, from proposition 1, 
\begin{equation}
\mathbf{G}+\mathbf{G}^{\dag }+\mathbf{G}^{\dag }\mathbf{PG}=\mathbf{0},\quad 
\mathbf{G}+\mathbf{G}^{\dag }+\mathbf{GPG}^{\dag }=\mathbf{0}.
\end{equation}
It is relatively easy to see that the It\={o} coefficients then take the
general form 
\begin{eqnarray}
\mathbf{PGP} &=&\mathbf{W}-\mathbf{P},  \notag \\
\mathbf{QGP} &=&-\left( \mathbf{PGQ}\right) ^{\dag }\mathbf{W},  \notag \\
\mathbf{QGQ} &=&-\frac{1}{2}\left( \mathbf{PGQ}\right) ^{\dag }\left( 
\mathbf{PGQ}\right) -\mathrm{i}\left( 
\begin{tabular}{l|l}
$H$ & $0$ \\ \hline
$0$ & $\mathsf{0}$%
\end{tabular}
\right) ,
\end{eqnarray}
where $\mathbf{W}^{\dag }\mathbf{W}=\mathbf{P}=\mathbf{WW}^{\dag }$ (i.e.,
the restriction of $\mathbf{W}$ to $\frak{h}\otimes \mathbb{C}^{N}$ is
unitary) and $H$ is self-adjoint on $\frak{h}$.

More explicitly, we may set 
\begin{equation*}
\mathbf{W}=\left( 
\begin{tabular}{l|l}
$0$ & $0$ \\ \hline
$0$ & $\mathsf{W}=\left\{ W_{ij}\right\} $%
\end{tabular}
\right) ,\quad \mathbf{PGQ}=\left( 
\begin{tabular}{l|l}
$0$ & $0,0,\cdots $ \\ \hline
$L_{1}$ &  \\ 
$L_{2}$ & \multicolumn{1}{|c}{$\mathsf{0}$} \\ 
$\vdots $ & 
\end{tabular}
\right)
\end{equation*}
where $W_{ij},L_{i}$ and $H$ are the operators on $\frak{h}$ introduced in
proposition 1.

We should remark that the restriction to a finite number $N$ of channels is
not essential and that the unitary process exists under certain conditions
on the boundedness of $\mathbf{G}$\ as a matrix operator \cite{Partha}.

\section{Quantum Stratonovich Calculus}

We wish to write the quantum It\={o} formula in the form

\begin{equation}
d\left( X_{t}Y_{t}\right) =\left( dX_{t}\right) \circ Y_{t}+X_{t}\circ
\left( dY_{t}\right) .
\end{equation}
This can be achieved by formally defining 
\begin{eqnarray}
\left( dX_{t}\right) \circ Y_{t} &:&=\left( dX_{t}\right)
Y_{t}+X_{t}^{\alpha \mu }V_{t}^{\mu \nu }Y_{t}^{\nu \beta }\,dA_{t}^{\alpha
\beta },  \notag \\
X_{t}\circ \left( dY_{t}\right) &:&=X_{t}\left( dY_{t}\right) +X_{t}^{\alpha
\mu }\left( V_{t}^{\nu \mu }\right) ^{\dag }Y_{t}^{\nu \beta
}\,dA_{t}^{\alpha \beta },
\end{eqnarray}
where the $\left\{ V_{t}^{\alpha \beta }\right\} $ may in general be taken
as adapted processes, however, we shall take them to be just scalar
coefficients. It follows that $\left[ \left( dX_{t}\right) \circ Y_{t}\right]
^{\dag }=Y_{t}^{\dag }\circ dX_{t}^{\dag }$, and we recover the It\={o}
formula provided we have the condition $V^{\mu \nu }+\left( V^{\nu \mu
}\right) ^{\ast }=P^{\mu \nu }$. The simplest solution possible is to take $%
V^{\mu \nu }=\frac{1}{2}P^{\mu \nu }$ and this corresponds algebraically to
the traditional Stratonovich definition of a differential. The general
solution however takes the form $V^{\mu \nu }=\frac{1}{2}P^{\mu \nu }+%
\mathrm{i}Z^{\mu \nu }$, where the constants $\left\{ Z^{\mu \nu }\right\} $
satisfy $\left( Z^{\mu \nu }\right) ^{\ast }=Z^{\nu \mu }$. The appearance
of these constants is similar to the ambiguity in the Tomita-Takesaki
theory, and we refer to them as a gauge freedom. We shall identify the $%
V^{ij}$ with the constants $\kappa _{ij}$ occurring in the approximation
scheme. Let us take $\mathbf{V}\equiv \left\{ V^{\alpha \beta }\right\} $ to
be the family of constants, then the requirement is $\mathbf{V}+\mathbf{V}%
^{\dag }=\mathbf{P}$ with general solution $\mathbf{V}\equiv \frac{1}{2}%
\mathbf{P}+\mathrm{i}\mathbf{Z}$ where $\mathbf{Z}^{\dag }=\mathbf{Z}$. We
shall take the $\left\{ Z^{\alpha \beta }\right\} $ to be scalar constants
and set $Z^{00}=Z^{i0}=Z^{0j}=0$. This implies that $\mathbf{Z}=\mathbf{ZP}=%
\mathbf{PZ}$ and so $\mathbf{V}=\mathbf{VP}=\mathbf{PV:}$

\begin{equation*}
\mathbf{V}=\frac{1}{2}\mathbf{P}+\mathrm{i}\mathbf{Z}\equiv \left( 
\begin{tabular}{l|l}
$0$ & $0$ \\ \hline
$0$ & $\mathsf{V}$%
\end{tabular}
\right) .
\end{equation*}
where $\mathsf{V}=\frac{1}{2}+\mathrm{i}\mathsf{Z}$ with $\mathsf{Z}^{\dag }=%
\mathsf{Z}$. Note that $\mathsf{V}$ is a normal operator with $\mathsf{VV}%
^{\dag }=\mathsf{V}^{\dag }\mathsf{V}=4^{-1}+\mathsf{Z}^{2}$.

It should be pointed out that we have the relation 
\begin{equation*}
dJ_{t}\left( XY\right) =\left( dJ_{t}\left( X\right) \right) \circ
J_{t}\left( Y\right) +J_{t}\left( X\right) \circ \left( dJ_{t}\left(
Y\right) \right)
\end{equation*}
which we can get either from taking differentials of the homomorphic
property $J_{t}\left( XY\right) =J_{t}\left( X\right) J_{t}\left( Y\right) $%
, or explicitly by noting that $\left( dJ_{t}\left( X\right) \right) \circ
J_{t}\left( Y\right) \equiv J_{t}(\mathcal{L}_{\alpha \beta }\left( X\right)
Y+\mathcal{L}_{\alpha \mu }\left( X\right) V^{\mu \nu }\mathcal{L}_{n\beta
}\left( Y\right) )dA_{t}^{\alpha \beta }$, etc., and using proposition 3.

\subsection{Stratonovich-Dyson Time Ordered Exponentials}

Let us now suppose that $U_{\cdot }$ is simultaneously the solution to the
It\={o} QSDE $dU=\left( dG\right) U$, with $dG=G_{\alpha \beta }dA^{\alpha
\beta }$ as before, and a Stratonovich QSDE (for a fixed gauge!)

\begin{equation}
dU_{t}=\left( dG_{0}\left( t\right) \right) \circ U_{t},\quad U_{0}=1,
\end{equation}
with $dG_{0}=G_{\alpha \beta }^{0}dA^{\alpha \beta }$. In such cases, we may
write $U_{\cdot }$ as the \textit{Dyson-Stratonovich time-ordered exponential%
} $U_{t}=\mathbf{\vec{T}}_{SD}\left\{ \exp \int_{0}^{t}G_{\alpha \beta
}^{0}dA^{\alpha \beta }\right\} $ and shall refer to $\mathbf{G}_{0}\equiv
\left\{ G_{\alpha \beta }^{0}\right\} $\ as the matrix of \textit{%
Stratonovich coefficients}.

Self-consistency requires that $dU=\left( dG^{0}\right) \circ U=\left(
dG\right) U$ and so we should have that $dU=\left( dG^{0}\right) U+G_{\alpha
\mu }^{0}V^{\mu \nu }G_{\nu \beta }\,UdA^{\alpha \beta }=\left( G_{\alpha
\beta }^{0}+G_{\alpha \mu }^{0}V^{\mu \nu }G_{\nu \beta }\right)
U\,dA^{\alpha \beta }$. This means that the It\={o} coefficients $\mathbf{G}%
=\left\{ G_{\alpha \beta }\right\} $ are related to the Stratonovich
coefficients $\mathbf{G}_{0}=\left\{ G_{\alpha \beta }^{0}\right\} $ by 
\begin{equation}
\mathbf{G}=\mathbf{G}_{0}+\mathbf{G}_{0}\mathbf{VG}\text{.}
\end{equation}

As we shall see, so long as $\mathbf{1}+\mathbf{VGP}$ is invertible, we may
solve for $\mathbf{G}_{0}$\ in terms of $\mathbf{G}$. Similarly,
invertibility of $\mathbf{1}-\mathbf{PG}_{0}\mathbf{V}$ implies that we may
write $\mathbf{G}$ in terms of $\mathbf{G}_{0}$. It might be remarked that
the relation $\left( 14\right) $ also applies if we consider matrices $%
\mathbf{G},\mathbf{G}_{0}$ of adapted processes.

What is rather astonishing is that relation $\left( 14\right) $ is precisely
of the form relating free and perturbed Green's functions. Let us recall
briefly that if $H=H_{0}+V$ is a Hamiltonian considered as a perturbation of
the free Hamiltonian $H_{0}$ then the resolvent operator $\mathcal{G}\left(
z\right) =\left( z-H\right) ^{-1}$ is related to the free resolvent $%
\mathcal{G}_{0}\left( z\right) =\left( z-H_{0}\right) ^{-1}$ by the
algebraic identity 
\begin{equation}
\mathcal{G}=\mathcal{G}_{0}+\mathcal{G}_{0}V\mathcal{G}
\end{equation}
for all $z$ outside of the spectra of $H$ and $H_{0}$. The identity may be
rewritten as $\mathcal{G}=\left( 1-\mathcal{G}_{0}V\right) ^{-1}\mathcal{G}%
_{0}$ and iterated to give the formal expansion $\mathcal{G}=\mathcal{G}_{0}+%
\mathcal{G}_{0}V\mathcal{G}_{0}+\mathcal{G}_{0}V\mathcal{G}_{0}V\mathcal{G}%
_{0}+\cdots $ which, when convergent, is the Neumann series. The details of
the actually scattering are contained in the operator $\mathcal{T}:=V+V%
\mathcal{G}V$ and we have the identity $\mathcal{G}=\mathcal{G}_{0}+\mathcal{%
G}_{0}\mathcal{TG}_{0}$.

\subsection{The $\mathbf{T}$-matrix}

We now exploit the similarity between $\left( 14\right) $ and $\left(
15\right) $. We begin by introducing the operator 
\begin{equation}
\mathbf{T}:=\mathbf{V}+\mathbf{VGV}\equiv \left( 
\begin{tabular}{l|l}
$0$ & $0$ \\ \hline
$0$ & $\mathsf{T}$%
\end{tabular}
\right) \mathbf{.}
\end{equation}
Assuming that $1-\mathbf{VG}_{0}\mathbf{P}$ is again invertible, we obtain
the following identities 
\begin{align}
\mathbf{T}& =\frac{1}{\mathbf{1}-\mathbf{VG}_{0}\mathbf{P}}\mathbf{V}, \\
\mathbf{G}_{0}\mathbf{T}& =\mathbf{GV}, \\
\mathbf{TG}_{0}& =\mathbf{VG}, \\
\mathbf{G}& =\mathbf{G}_{0}+\mathbf{GVG}_{0}, \\
\mathbf{G}& =\mathbf{G}_{0}+\mathbf{G}_{0}\mathbf{TG}_{0}.
\end{align}

The proof of $\left( 17\right) $ comes from writing 
\begin{equation*}
\mathbf{T}=\mathbf{V}+\mathbf{T}\left( \mathbf{G}_{0}+\mathbf{G}_{0}\mathbf{%
TG}\right) \mathbf{V}=\mathbf{V}+\mathbf{VG}_{0}\mathbf{T}
\end{equation*}
so that $\left( \mathbf{1}-\mathbf{VG}_{0}\mathbf{P}\right) \mathbf{T}=%
\mathbf{V}$. The remaining identities are just precise analogues of
well-known relations for resolvent operators \cite{Economou}.

\bigskip

Combining $\left( 17\right) $ and $\left( 21\right) $ we see that $\mathbf{G}
$ can be expressed in terms of $\mathbf{G}_{0}$ as 
\begin{equation}
\mathbf{G=G}_{0}+\mathbf{G}_{0}\left( \mathbf{1}-\mathbf{PVG}_{0}\mathbf{P}%
\right) ^{-1}\mathbf{VG}_{0}.
\end{equation}
In particular, we see that $\mathbf{G}$ is bounded. We may then invert to
get 
\begin{equation}
\mathbf{G}_{0}\mathbf{=G}-\mathbf{G}\left( \mathbf{1}+\mathbf{PVGP}\right)
^{-1}\mathbf{VG}.
\end{equation}

The equations $\left( 22\right) $ and $\left( 23\right) $ reveal a
remarkable duality between the It\={o} and Stratonovich coefficients. (Of
course this just means that we may view either as a ``perturbation'' of the
other!)

If $\mathbf{PVG}_{0}\mathbf{P}$ is a strict contraction, then we may develop
a Neumann series expansion $\mathbf{G=G}_{0}+\mathbf{G}_{0}\mathbf{VG}_{0}+%
\mathbf{G}_{0}\mathbf{VG}_{0}\mathbf{VG}_{0}+\cdots =\mathbf{G}%
_{0}\sum_{n=0}^{\infty }\left( \mathbf{VG}_{0}\right) ^{n}$.

It is convenient to introduce a related matrix 
\begin{equation}
\mathbf{F}=\mathbf{1}+\mathbf{TG}_{0}=\mathbf{1}+\mathbf{VG}
\end{equation}
so that $\mathbf{G=G}_{0}\mathbf{F}$.

\subsection{An ``Optical Theorem''}

Let us next suppose that the Stratonovich coefficients take the \textit{%
Hamiltonian} form 
\begin{equation}
\mathbf{G}_{0}\mathbf{=}-\mathrm{i}\mathbf{E}
\end{equation}
where $\mathbf{E}$ is a bounded, self-adjoint operator on $\frak{h}\otimes 
\mathbb{C}^{N+1}$. We then have the relation $\mathbf{G}_{0}^{\dag }=-%
\mathbf{G}_{0}$ and set 
\begin{equation*}
\mathbf{PEP}=\left( 
\begin{tabular}{l|l}
$0$ & $0$ \\ \hline
$0$ & $\mathsf{E}$%
\end{tabular}
\right)
\end{equation*}
so that $\mathsf{E}$ is self-adjoint on $\frak{h}\otimes \mathbb{C}^{N}$.
When our invertibility condition is met, it is easy to see that matrix $%
\mathsf{T}$ exists and can be written as 
\begin{equation}
\mathsf{T}=\left[ 1+\mathrm{i}\mathsf{VE}\right] ^{-1}\mathsf{V}\equiv \frac{%
1}{\mathsf{V}^{-1}+\mathrm{i}\mathsf{E}}.
\end{equation}

(In the special case where $\mathsf{Z}=0$, the self-adjointness of $\mathsf{E%
}$ ensures that $\mathsf{V}=1+\mathrm{i}\frac{1}{2}\mathsf{E}$ is invertible
by von Neumann's theorem \cite{Reed Simon}. Therefore the existence of
matrix $\mathsf{T}$ is guaranteed. More generally, so long as the value $1$
lies in the resolvent set of $\mathsf{ZE}$, this theorem implies the
existence of $\mathsf{T}$.)

The related matrix $\mathbf{F}$ then takes the form 
\begin{equation*}
\mathbf{F}=\left( 
\begin{tabular}{l|l}
$1$ & $0,0,\cdots $ \\ \hline
$-\mathrm{i}T_{1j}E_{j0}$ &  \\ 
$-\mathrm{i}T_{2j}E_{j0}$ & \multicolumn{1}{|c}{$\mathsf{F}$} \\ 
$\vdots $ & 
\end{tabular}
\right)
\end{equation*}
with $\mathsf{F}=\mathsf{1}-\mathrm{i}\mathsf{TE}=\mathsf{T}\mathsf{V}^{-1}$%
, so that $\mathsf{F}\equiv \left[ 1+\mathrm{i}\mathsf{VE}\right] ^{-1}$.

\bigskip

\noindent \textbf{Lemma 1 (``Optical Theorem''): }$\func{Re}\mathsf{T}\geq 0$
\emph{and in particular }$\mathsf{T}$\emph{\ satisfies the identity} 
\begin{equation}
\mathsf{T}+\mathsf{T}^{\dag }=\mathsf{FF}^{\dag }=\mathsf{F}^{\dag }\mathsf{F%
}.
\end{equation}
.

\begin{proof}
We have that$\mathsf{T}+\mathsf{T}^{\dag }$ may be written as 
\begin{equation*}
\mathsf{FV}+\mathsf{V}^{\dag }\mathsf{F}^{\dag }=\mathsf{F}\left[ \mathsf{V}%
\left( \mathsf{1}-\mathrm{i}\mathsf{EV}^{\dag }\right) +\left( \mathsf{1}+%
\mathrm{i}\mathsf{VE}\right) \mathsf{V}^{\dag }\right] \mathsf{F}^{\dag }=%
\mathsf{F}\left[ \mathsf{V}+\mathsf{V}^{\dag }\right] \mathsf{F}^{\dag }=%
\mathsf{FF}^{\dag }.
\end{equation*}
It is then relatively straightforward to show that 
\begin{eqnarray*}
\mathsf{FF}^{\dag } &=&\frac{1}{\left( 1-\mathrm{i}\mathsf{V}^{\dag }\mathsf{%
E}\right) \left( 1+\mathrm{i}\mathsf{EV}\right) }=\left[ 1-\mathsf{EZ}-%
\mathsf{ZE}+\mathsf{E\mathsf{V}}\left( \mathsf{\mathsf{V}}^{\dag }\right) 
\mathsf{E}\right] ^{-1} \\
&=&\frac{1}{\left( 1+\mathrm{i}\mathsf{EV}\right) \left( 1-\mathrm{i}\mathsf{%
V}^{\dag }\mathsf{E}\right) }=\mathsf{F}^{\dag }\mathsf{F}.
\end{eqnarray*}
\end{proof}

A similar calculation shows that $\func{Im}\mathsf{T}=-\mathsf{TET}^{\dag }.$

\subsection{Unitarity}

We now wish to show that the choice of Hamiltonian Stratonovich
coefficients\ naturally leads to unitary processes.

\bigskip

\noindent \textbf{Lemma 2: }\emph{Let} $\mathbf{G}_{0}^{\dag }=-\mathbf{G}%
_{0}$ \emph{be bounded with} $\mathbf{1}-\mathbf{VG}_{0}\mathbf{P}$ \emph{%
invertible, and set }$G^{0}\left( t\right) =\int_{0}^{t}G_{\alpha \beta
}^{0}dA^{\alpha \beta }$\emph{. Then the solution to the Stratonovich QSDE }$%
dU=\left( dG^{0}\right) \circ U$\emph{, }$U_{0}=1$\emph{\ will be unitary.}

\bigskip

\begin{proof}
This fact is an immediate consequence of the optical theorem $\left(
27\right) $. To establish the isometric property for the It\={o}
coefficients, first observe that $\mathbf{G}+\mathbf{G}^{\dag }=-\mathbf{G}%
_{0}\left( \mathbf{T}+\mathbf{T}^{\dag }\right) \mathbf{G}_{0}$ while 
\begin{equation*}
\mathbf{GPG}^{\dag }=\mathbf{G}_{0}\mathbf{FVF}^{\dag }\mathbf{G}_{0}^{\dag
}=\mathbf{G}_{0}\left( 
\begin{tabular}{l|l}
$0$ & $0$ \\ \hline
$0$ & $\mathsf{FF}^{\dag }$%
\end{tabular}
\right) \mathbf{G}_{0}^{\dag }=-\mathbf{G}_{0}\left( \mathbf{T}+\mathbf{T}%
^{\dag }\right) \mathbf{G}_{0}.
\end{equation*}
The isometry condition in $\left( 9\right) $ then follows from the first
part of $\left( 27\right) $. The co-isometric property likewise follows from
the second part.
\end{proof}

\subsection{Changing Gauge}

Let $\mathbf{G}$ be a fixed It\={o} coefficient matrix related to the
Stratonovich coefficient matrices $\mathbf{G}_{0}^{\left( a\right) }$ and $%
\mathbf{G}_{0}^{\left( a\right) }$ with gauges $\mathbf{Z}^{\left( a\right)
} $ and $\mathbf{Z}^{\left( a\right) }$, respectively. The two matrices will
then be related by the perturbative formula 
\begin{equation*}
\mathbf{G}_{0}^{\left( a\right) }=\mathbf{G}_{0}^{\left( a\right) }+\mathrm{i%
}\mathbf{G}_{0}^{\left( a\right) }\left( \mathbf{Z}^{\left( a\right) }-%
\mathbf{Z}^{a}\right) \mathbf{G}_{0}^{\left( a\right) }.
\end{equation*}
In particular, we can relate $\mathbf{G}_{0}^{\left( a\right) }$ for a
non-zero gauge to the symmetric (gauge zero) form $\mathbf{G}_{0}^{\left(
a\right) }$. As we shall see, the gauge $\mathbf{Z}$ has the physically
interpretation as the imaginary part in the complex damping and in many
applications this may be small \cite{C.W. Gardiner}.

\section{Wick Ordering Rule}

Let $X_{\cdot }$ and $Y_{\cdot }$ be quantum stochastic integrals with
adapted integrands as before. In terms of our notation involving the $%
a_{\alpha }^{\dag }\left( t\right) ,a_{\beta }\left( t\right) $, the product 
$X_{t}Y_{t}=\int_{0}^{t}ds_{1}\int_{0}^{t}ds_{2}\,a_{\alpha }^{\dag }\left(
s_{1}\right) X_{s_{1}}^{\alpha \beta }a_{\beta }\left( s_{1}\right) a_{\mu
}^{\dag }\left( s_{2}\right) Y_{s_{2}}^{\mu \nu }a_{\nu }\left( s_{2}\right) 
$ is not immediately interpreted as an iterated integral since it is out of
Wick order. However, the rule for achieving this is formally equivalent to
the kinematic relations 
\begin{equation}
\left[ a_{\alpha }\left( t\right) ,Y_{s}\right] =\left\{ 
\begin{array}{cc}
P^{\alpha \mu }Y_{t}^{\mu \nu }a_{\nu }\left( t\right) , & t<s; \\ 
V^{\alpha \mu }Y_{t}^{\mu \nu }a_{\nu }\left( t\right) , & t=s; \\ 
0, & t>s;
\end{array}
\right.
\end{equation}
under the integral sign, along with its adjoint $\left[ X_{t},a_{\beta
}^{\dag }\left( t\right) \right] =\left[ a_{\beta }\left( t\right)
,X_{s}^{\dag }\right] ^{\dag }$. These relations can be viewed as the formal
commutation relations $\left[ a_{\alpha }\left( t\right) ,a_{\beta }^{\dag
}\left( s\right) \right] =V^{\alpha \beta }\delta _{+}\left( t-s\right)
+\left( V^{\beta \alpha }\right) ^{\ast }\delta _{-}\left( t-s\right) $ at
work, where the $\delta _{\pm }$ are one-sided delta-functions: $\int \delta
_{\pm }\left( f\right) f\left( t\right) =f\left( 0^{\pm }\right) $, see e.g. 
\cite{GTMP},\cite{WvW}. It is possible to interpret the $a_{\alpha
}^{\#}\left( t\right) $ as quantum white noise operators, but we do not
stress this point further here. At this stage, we could switch to a fluxion
notation such as $\dot{X}_{t}=\frac{dX_{t}}{dt}=a_{\alpha }^{\dag }\left(
t\right) X_{t}^{\alpha \beta }a_{\beta }\left( t\right) $, etc., and write
the It\={o} formula as $\frac{d}{dt}\left( X_{t}Y_{t}\right) =\dot{X}%
_{t}\circ Y_{t}+X_{t}\circ \dot{Y}_{t}$ with the convention that 
\begin{equation*}
\dot{X}_{t}\circ Y_{t}\equiv a_{\alpha }^{\dag }\left( t\right)
X_{t}^{\alpha \beta }a_{\beta }\left( t\right) Y_{t}=a_{\alpha }^{\dag
}\left( t\right) X_{t}^{\alpha \beta }Y_{t}a_{\beta }\left( t\right)
+a_{\alpha }^{\dag }\left( t\right) X_{t}^{\alpha \mu }V^{\mu \nu
}Y_{t}^{\nu \beta }a_{\beta }\left( t\right) .
\end{equation*}

In particular, we have the following interpretation of the results of the
previous section: The equation $\dot{U}_{t}=\dot{G}_{0}\left( t\right) U_{t}$
with $\dot{G}_{0}\left( t\right) =a_{\alpha }^{\dag }\left( t\right)
G_{\alpha \beta }^{0}a_{\beta }\left( t\right) $ is out of Wick order, but
can be put to Wick order as $\dot{U}_{t}=a_{\alpha }^{\dag }\left( t\right)
G_{\alpha \beta }U_{t}a_{\beta }\left( t\right) $. The relation $\left[
a_{\alpha }\left( t\right) ,U_{t}\right] =V^{\alpha \mu }G^{\mu \nu
}U_{t}a_{\nu }\left( t\right) $ implies that 
\begin{equation}
a_{\alpha }\left( t\right) U_{t}=\left( \delta _{\alpha \beta }+V^{\alpha
\mu }G_{\mu \beta }\right) U_{t}a_{\beta }\left( t\right) =F_{\alpha \beta
}U_{t}a_{\beta }\left( t\right) ,
\end{equation}
where $\left\{ F_{\alpha \beta }\right\} $ are the components of the matrix $%
\mathbf{F}$ introduced in $\left( 24\right) $.

\bigskip

The QSDE for the unitary $U_{\cdot }$, with $G_{\alpha \beta }^{0}=-\mathrm{i%
}E_{\alpha \beta }$, will then be 
\begin{equation}
\dot{U}_{t}=-\mathrm{i}a_{\alpha }^{\dag }\left( t\right) E_{\alpha \beta
}a_{\beta }\left( t\right) U_{t}=-\mathrm{i}a_{\alpha }^{\dag }\left(
t\right) E_{\alpha \beta }F_{\beta \nu }U_{t}a_{\nu }\left( t\right)
\end{equation}
and likewise the QSDE for the flow will be 
\begin{eqnarray}
\frac{d}{dt}J_{t}\left( X\right) &=&\dot{U}_{t}^{\dag }\left( X\right)
U_{t}+U_{t}^{\dag }\left( X\right) \dot{U}_{t}  \notag \\
&=&-\mathrm{i}U_{t}^{\dag }a_{\alpha }^{\dag }\left( t\right) \left[
X,E_{\alpha \beta }\right] a_{\beta }\left( t\right) U_{t}  \notag \\
&=&-\mathrm{i}a_{\mu }^{\dag }\left( t\right) U_{t}^{\dag }F_{\alpha \mu
}^{\dag }\left[ X,E_{\alpha \beta }\right] F_{\beta \nu }U_{t}a_{\nu }\left(
t\right) .
\end{eqnarray}
Comparison with proposition 2 suggest that $\mathcal{L}_{\mu \nu }\left(
X\right) =F_{\alpha \mu }^{\dag }\left[ X,E_{\alpha \beta }\right] F_{\beta
\nu }$. As this is an entirely new relation, we give an independent
derivation in appendix A using only the It\={o} calculus.

\bigskip

\noindent \textbf{Lemma 3:} \emph{Under the conventions and notations of the
previous sections, the Evans-Hudson maps take the form} 
\begin{equation}
\mathcal{L}_{\alpha \beta }\left( X\right) =-\mathrm{i}F_{\mu \alpha }^{\dag
}\left[ X,E_{\mu \nu }\right] F_{\nu \beta }.
\end{equation}

\section{Quantum Wong-Zakai Theorem}

The following is the multi-dimensional version of a result first established
in \cite{GQCLT}.

\noindent \textbf{Theorem:} \emph{\ Let }$a_{\alpha }^{\#}\left( t,\lambda
\right) $\emph{, }$\alpha =0,1,\cdots ,N,$\emph{\ be continuous in }$t$\emph{%
\ creation / annihilation fields for each }$\lambda >0$\emph{\ with }$%
A_{t}^{\alpha \beta }\left( \lambda \right) =\int_{0}^{t}a_{\alpha }^{\dag
}\left( s,\lambda \right) a_{\beta }\left( s,\lambda \right) ds$\emph{\
approximating fundamental quantum stochastic processes with internal space }$%
\mathbb{C}^{N}$\emph{\ as before, with fixed gauge matrix }$\mathbf{V}%
=\left\{ V^{\alpha \beta }\right\} $\emph{. If }$\Upsilon _{t}^{\left(
\lambda \right) }=E_{\alpha \beta }\otimes a_{\alpha }^{\dag }\left(
t,\lambda \right) a_{\beta }\left( t,\lambda \right) $\emph{\ with }$%
E_{\alpha \beta }^{\dag }=E_{\beta \alpha }$\emph{\ bounded operators on a
fixed Hilbert space }$\frak{h}$\emph{\ such that }$\mathsf{VE}$\emph{\ is a
strict contraction, then the unitary family }$U_{\cdot }^{\left( \lambda
\right) }$\emph{\ and the Heisenberg dynamical map }$J_{\cdot }^{\left(
\lambda \right) }\left( X\right) =U_{\cdot }^{\left( \lambda \right) \dag
}XU_{\cdot }^{\left( \lambda \right) }$\emph{, determined by the
Schr\"{o}dinger equation }$\dot{U}_{t}^{\left( \lambda \right) }=-\mathrm{i}%
\Upsilon _{t}^{\left( \lambda \right) }U_{t}^{\left( \lambda \right) },$%
\emph{\ }$U_{0}^{\left( \lambda \right) }=1$\emph{, converge in the sense of
weak matrix limits to the unitary quantum stochastic process }$U_{\cdot }$%
\emph{\ and corresponding quantum stochastic flow }$J_{\cdot }\left(
X\right) $\emph{. The limit process }$U_{\cdot }$\emph{\ is unitary adapted
and satisfies the Stratonovich QSDE }$dU_{t}=-\mathrm{i}E_{\alpha \beta
}U_{t}\circ dA_{t}^{\alpha \beta },$ $U_{0}=1,$\ \emph{with gauge determined
by }$\mathbf{V}=\left\{ V^{\alpha \beta }\right\} $\emph{.}

\bigskip

The condition $\left\| \mathsf{VE}\right\| <1$ gives convergence of the
Neumann series. It also implies that $\mathbf{1}+\mathbf{VEP}$ will be
invertible and therefore the Stratonovich QSDE makes sense. We will sketch
the proof of this theorem in the Appendix B. Provided that the strict
contractivity conditions hold, we could replace $\mathsf{E}$, and indeed $%
\mathsf{V}$, by suitably continuous adapted processes.

It might be remarked that there exists an analogue of this result using
Fermi fields in place of Bose fields \cite{GS05a}. The limit QSDE changes
insofar as the noises must now be Fermionic processes, however, the
coefficients are exactly as before.

\subsection{Examples}

\subsubsection{Classical Wong Zakai Theorem}

As a very special example of theorem, let us take the 1-dimensional case
with the pre-limit Hamiltonian $\Upsilon _{t}^{\left( \lambda \right) }$
determined by $E_{00}=H$, $E_{01}=E_{10}=R$ and $E_{11}=0$ with $\kappa =%
\frac{1}{2}$. Then the limit flow is characterized by the maps $\mathcal{L}%
_{11}\left( X\right) =0$, $\mathcal{L}_{10}\left( X\right) =\mathcal{L}%
_{01}\left( X\right) =-i\left[ X,R\right] $ and $\mathcal{L}_{00}\left(
X\right) =-i\left[ X,H\right] -\frac{1}{2}\left[ \left[ X,R\right] ,R\right] 
$. For the choices $H=\frac{1}{2}\left( pv\left( q\right) +v\left( q\right)
p\right) $ and $R=\frac{1}{2}\left( p\sigma \left( q\right) +\sigma \left(
q\right) p\right) $, where $v\left( \cdot \right) $ and $\sigma \left( \cdot
\right) $ are Lipschitz continuous with\emph{\ }$|v\left( x\right) |,|\sigma
\left( x\right) |<C\left( 1+|x|\right) $\emph{\ }for some constant $%
0<C<\infty $, and $q$ and $p$ are canonical position and momentum
observables, we have that $q_{t}:=J_{t}\left( q\right) $ satisfies the
essentially classical SDE $dq_{t}=\left[ v\left( q_{t}\right) +\sigma \left(
q_{t}\right) \sigma ^{\prime }\left( q_{t}\right) \right] dt+\sigma \left(
q_{t}\right) dQ_{t}$. or equivalently $dq_{t}=v\left( q_{t}\right) dt+\sigma
\left( q_{t}\right) \circ dQ_{t}$ where $Q_{t}=A_{t}^{10}+A_{t}^{01}$ is a
copy of the Wiener process.

As a result, the theorem reduces to a classical Wong-Zakai approximation
theorem which states that, since $Q_{t}^{\left( \lambda \right)
}=\int_{0}^{t}\left( a^{\dag }\left( s,\lambda \right) +a\left( s,\lambda
\right) \right) ds$ is an essentially classical stochastic process that is
differentiable in time $t$ and converges almost always uniformly on compact
time-intervals to a Wiener process $Q_{t}$, the solution to the random ODE $%
\dot{X}_{t}^{\left( \lambda \right) }=v\left( X_{t}^{\left( \lambda \right)
},t\right) +\sigma \left( X_{t}^{\left( \lambda \right) },t\right) \dot{Q}%
_{t}^{\left( \lambda \right) }$\emph{, }$X_{0}^{\left( \lambda \right)
}=x_{0}$ similarly converges to the diffusion process $X_{\cdot }$\
satisfying the Stratonovich SDE 
\begin{equation*}
dX_{t}=v\left( X_{t},t\right) dt+\sigma \left( X_{t},t\right) \circ
dQ_{t},\quad X_{0}=x_{0}.
\end{equation*}

\subsubsection{Quantum Diffusions}

Taking $\Upsilon _{t}^{\left( \lambda \right) }=R\otimes a^{\dag }\left(
t,\lambda \right) +R^{\dag }\otimes a\left( t,\lambda \right) +H$ leads to
the limit QSDE 
\begin{eqnarray*}
dU_{t} &=&-\mathrm{i}\left( R\otimes dA_{t}^{\dag }+R^{\dag }\otimes
dA_{t}+H\right) \circ U_{t} \\
&\equiv &-\mathrm{i}\left( R\otimes dA_{t}^{\dag }+R^{\dag }\otimes
dA_{t}+H\right) U_{t}-\kappa R^{\dag }RU_{t}dt.
\end{eqnarray*}
Note that $\func{Re}\kappa =\frac{1}{2}$ so that $dJ_{t}\left( X\right) =-%
\mathrm{i}J_{t}\left( \left[ X,R\right] \right) dA_{t}^{\dag }-\mathrm{i}%
J_{t}\left( \left[ X,R^{\dag }\right] \right) dA_{t}+J_{t}\left( \mathcal{L}%
\left( X\right) \right) dt$ where we set $A_{t}=A_{t}^{01}$, $A_{t}^{\dag
}=A_{t}^{10}$ and $\mathcal{L}\left( X\right) =\mathcal{L}_{00}\left(
X\right) =-\mathrm{i}\left[ X,H^{\prime }\right] +\frac{1}{2}\left[ R^{\dag
},X\right] R+\frac{1}{2}R^{\dag }\left[ X,R\right] $. The new operator $%
H^{\prime }$ is $H+\func{Im}\left\{ \kappa \right\} R^{\dag }R$ which
includes an energy shift coming from the complex damping $\kappa $. The
theorem then reduces to a long line of results dealing with broadband noise
limits, weak coupling limits, etc., in quantum physics.

It is straightforward to extend this to describe coherent states, thermal
states and squeezed states \cite{GGRS}.

\subsubsection{Counting Processes}

Let us consider the choice $\Upsilon _{t}^{\left( \lambda \right) }=E\otimes
\left\{ a\left( t,\lambda \right) +f\left( t\right) \right\} ^{\dag }\left\{
a\left( t,\lambda \right) +f\left( t\right) \right\} $. This is what we
could consider in the vacuum as an equivalent of studying $E\otimes a\left(
t,\lambda \right) ^{\dag }a\left( t,\lambda \right) $ in a coherent of
intensity $f$. Here we have $E_{\alpha \beta }=Ef_{\alpha }^{\ast }\left(
t\right) f_{\beta }\left( t\right) $ and the one dimensional form of the
theorem yields the It\={o} coefficients $G_{\alpha \beta }=-iE_{\alpha \beta
}-\kappa E_{\alpha 1}\frac{1}{1+\mathrm{i}\kappa E_{11}}E_{1\beta }$, which
in this case reduce to $G_{\alpha \beta }\left( t\right) =\frac{-\mathrm{i}E%
}{1+\mathrm{i}\kappa E}f_{\alpha }^{\ast }\left( t\right) f_{\beta }\left(
t\right) $. The It\={o} form of the limit QSDE is then 
\begin{equation*}
dU_{t}=\frac{-\mathrm{i}E}{1+\mathrm{i}\kappa E}dN_{t}\left( f\right) \,U_{t}
\end{equation*}
and we introduce $N_{t}\left( f\right) =\int_{0}^{t}\left(
dA_{s}^{11}+f\left( s\right) dA_{s}^{10}+f^{\ast }\left( s\right)
dA_{s}^{01}+|f\left( s\right) |^{2}ds\right) $ which is essentially
classical and corresponds to a time-inhomogeneous Poisson process with
instantaneous rate of change $\nu \left( t\right) =|f\left( t\right) |^{2}$.

\begin{center}
{\Large Appendix A (Proof of Lemma 3)}
\end{center}

We restrict to the $N=1$ case, as this is already somewhat involved. The
multi-dimensional case does not present any more technical difficulties.
Here the Evans-Hudson maps are given by 
\begin{eqnarray}
\mathcal{L}_{11}\left( X\right) &=&W^{\dag }XW-X;  \notag \\
\mathcal{L}_{10}\left( X\right) &=&W^{\dag }\left[ X,L\right] ;\;\mathcal{L}%
_{01}\left( X\right) =-\left[ X,L^{\dag }\right] W;  \notag \\
\mathcal{L}_{00}\left( X\right) &=&\frac{1}{2}\left[ L^{\dag },X\right] L+%
\frac{1}{2}L^{\dag }\left[ X,L\right] -\mathrm{i}\left[ X,H\right] .
\end{eqnarray}
where the operators $W,H,L$ have the explicit forms 
\begin{equation}
W=\frac{1-\mathrm{i}\kappa ^{\ast }E_{11}}{1+\mathrm{i}\kappa E_{11}},\;L=-%
\mathrm{i}\frac{1}{1+\mathrm{i}\kappa E_{11}}E_{10},\;H=E_{00}+E_{01}\func{Im%
}\left\{ \frac{\kappa }{1+\mathrm{i}\kappa E_{11}}\right\} E_{10}.
\end{equation}
The components of the matrix $F$ can also be written out in detail: 
\begin{equation}
F_{11}=\frac{1}{1+\mathrm{i}\kappa E_{11}},\;F_{10}=-\mathrm{i}\kappa \frac{1%
}{1+\mathrm{i}\kappa E_{11}}E_{10},\;F_{01}=0,\;F_{00}=1.
\end{equation}
We now check that the relation $\left( 32\right) $ is correct by direct
substitution.

For $\alpha =\beta =1$, we find after a little algebra that 
\begin{eqnarray*}
\mathcal{L}_{11}\left( X\right) &=&\frac{1+\mathrm{i}\kappa E_{11}}{1-%
\mathrm{i}\kappa ^{\ast }E_{11}}X\frac{1-\mathrm{i}\kappa ^{\ast }E_{11}}{1+%
\mathrm{i}\kappa E_{11}}-X \\
&=&-\mathrm{i}\frac{1}{1-\mathrm{i}\kappa ^{\ast }E_{11}}\left[ X,E_{11}%
\right] \frac{1}{1+\mathrm{i}\kappa E_{11}} \\
&=&-\mathrm{i}F_{\mu 1}^{\dag }\left[ X,E_{\mu \nu }\right] F_{\nu 1}.
\end{eqnarray*}

For $\alpha =1,\beta =0$, we have 
\begin{equation*}
\mathcal{L}_{10}\left( X\right) =\frac{1+\mathrm{i}\kappa E_{11}}{1-\mathrm{i%
}\kappa ^{\ast }E_{11}}\left[ X,-\mathrm{i}\frac{1}{1+\mathrm{i}\kappa E_{11}%
}E_{10}\right]
\end{equation*}
and to compute this we need the observation that 
\begin{equation}
\left[ X,\frac{1}{1+\mathrm{i}\kappa E_{11}}\right] =-\mathrm{i}\kappa \frac{%
1}{1+\mathrm{i}\kappa E_{11}}\left[ X,E_{11}\right] \frac{1}{1+\mathrm{i}%
\kappa E_{11}}
\end{equation}
to write 
\begin{eqnarray*}
\mathcal{L}_{10}\left( X\right) &=&i\kappa \frac{1}{1-\mathrm{i}\kappa
^{\ast }E_{11}}\left[ X,E_{11}\right] \frac{1}{1+\mathrm{i}\kappa E_{11}}-%
\mathrm{i}\frac{1}{1-\mathrm{i}\kappa ^{\ast }E_{11}}\left[ X,E_{10}\right]
\\
&=&-\mathrm{i}F_{\mu 1}^{\dag }\left[ X,E_{\mu \nu }\right] F_{\nu 0}.
\end{eqnarray*}
As we have $\mathcal{L}_{\alpha \beta }\left( X\right) ^{\dag }=\mathcal{L}%
_{\beta \alpha }\left( X^{\dag }\right) $, this gives the $\mathcal{L}%
_{01}\left( X\right) $ result also.

The final case is the Lindbladian map $\mathcal{L}_{00}$. Substituting in
gives 
\begin{gather*}
\mathcal{L}_{00}\left( X\right) =-\mathrm{i}\left[ X,E_{00}\right] -\mathrm{i%
}\left[ X,E_{01}\func{Im}\left\{ T\right\} E_{10}\right] \\
+\frac{1}{2}\left[ E_{01}\frac{1}{1-\mathrm{i}\kappa ^{\ast }E_{11}},X\right]
\frac{1}{1+\mathrm{i}\kappa E_{11}}E_{10}+\frac{1}{2}E_{01}\frac{1}{1-%
\mathrm{i}\kappa ^{\ast }E_{11}}\left[ X,\frac{1}{1+\mathrm{i}\kappa E_{11}}%
E_{10}\right]
\end{gather*}
and again we use $\left( 36\right) $ along with the observation that 
\begin{equation*}
\func{Re}T=\frac{1}{2}\frac{1}{1-\mathrm{i}\kappa ^{\ast }E_{11}}\frac{1}{1+%
\mathrm{i}\kappa E_{11}}
\end{equation*}
to obtain 
\begin{multline*}
\mathcal{L}_{00}\left( X\right) =-\mathrm{i}\left[ X,E_{00}\right] -\left[
X,E_{01}\right] TE_{10}+E_{01}T^{\dag }\left[ X,E_{10}\right] \\
-\mathrm{i}E_{01}\left[ X,\func{Im}T\right] E_{10}-\mathrm{i}E_{01}T^{\dag }%
\left[ X,E_{11}\right] \func{Re}\left\{ T\right\} E_{10}-\mathrm{i}E_{01}%
\func{Re}\left\{ T^{\dag }\right\} \left[ X,E_{11}\right] TE_{10}.
\end{multline*}
The first three terms are $-\mathrm{i}F_{\alpha 0}^{\dag }\left[ X,E_{\alpha
\beta }\right] F_{\beta 0}$, \ for $\left( \alpha ,\beta \right) =\left(
0,0\right) ,\left( 0,1\right) $ and $\left( 1,0\right) $\ respectively. To
tidy up the last term, we note the identities 
\begin{equation*}
\left[ X,T\right] =-\mathrm{i}T\left[ X,E_{11}\right] T,\;\left[ X,T^{\dag }%
\right] =+\mathrm{i}T^{\dag }\left[ X,E_{11}\right] T^{\dag }
\end{equation*}
so that 
\begin{eqnarray*}
&&\left[ X,\func{Im}T\right] +T^{\dag }\left[ X,E_{11}\right] \func{Re}T+%
\func{Re}T\,\left[ X,E_{11}\right] T \\
&=&-\frac{1}{2}T\left[ X,E_{11}\right] T-\frac{1}{2}T^{\dag }\left[ X,E_{11}%
\right] T+T^{\dag }\left[ X,E_{11}\right] \func{Re}T+\func{Re}T\,\left[
X,E_{11}\right] T \\
&\equiv &T^{\dag }\left[ X,E_{11}\right] T
\end{eqnarray*}
and therefore the last term is $-\mathrm{i}E_{01}T^{\dag }\left[ X,E_{11}%
\right] TE_{10}\equiv $ $-\mathrm{i}F_{10}^{\dag }\left[ X,E_{10}\right]
F_{10}$. This gives the desired form.

\begin{center}
{\Large Appendix B (Proof of the Theorem)}
\end{center}

We now sketch briefly the proof of limit using diagrams. Essentially, this
is a form of the van Kampen cumulant expansion \cite{C.W. Gardiner} which
can be described explicitly. The Heisenberg evolution limit is similar
though more involved \cite{GQCLT}.

\textbf{Step 1: Wick Ordering the Dyson Series}

For finite $\lambda $, the mapping $t\mapsto U_{t}^{\left( \lambda \right) }$
is differentiable, and we may formally expand as the Dyson series $%
U_{t}^{\left( \lambda \right) }=\sum_{n\geq 0}\left( -\mathrm{i}\right)
^{n}\int_{\Delta _{n}\left( t\right) }\Upsilon _{t_{n}}^{\left( \lambda
\right) }\cdots \Upsilon _{t_{1}}^{\left( \lambda \right) }$. Here $\Delta
_{n}\left( t\right) $ is the simplex consisting of multi-times $\left(
t_{n},\cdots ,t_{1}\right) $ with $t\geq t_{n}\geq \cdots \geq t_{1}\geq 0$.

It is convenient to put the Dyson series to Wick order using the commutation
relations. The most convenient way to describe this is to expand in terms of
diagrammatic series and we borrow some standard techniques from field
theory. To this end, we introduce four vertices corresponding to the four
components of $\Upsilon _{t}^{\left( \lambda \right) }$:

\bigskip

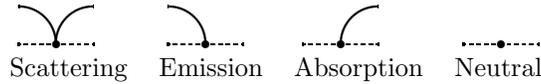
\begin{center}
\begin{tabular}{llll}
%
%
%
%
\setlength{\unitlength}{.1cm}
\begin{picture}(10,5)
\label{picdo}

\put(0,0){\dashbox{0.5}(10,0){ }}

\thicklines

\put(5,0){\circle*{1}}
\put(0,0){\oval(10,10)[tr]}
\put(10,0){\oval(10,10)[tl]}

\end{picture}
%
& 
%
%
%
\setlength{\unitlength}{.1cm}
\begin{picture}(10,5)
\label{pical}

\put(0,0){\dashbox{0.5}(10,0){ }}

\thicklines
\put(5,0){\circle*{1}}
\put(0,0){\oval(10,10)[tr]}

\end{picture}
%
& 
%
%
%
%
\setlength{\unitlength}{.1cm}
\begin{picture}(10,5)
\label{picb}

\put(0,0){\dashbox{0.5}(10,0){ }}

\thicklines

\put(5,0){\circle*{1}}
\put(10,0){\oval(10,10)[tl]}

\end{picture}
%
& 
%
%
%
%
%
\setlength{\unitlength}{.1cm}
\begin{picture}(10,5)
\label{picd}

\put(0,0){\dashbox{0.5}(10,0){ }}

\thicklines

\put(5,0){\circle*{1}}

\end{picture}
%
\\ 
Scattering & Emission & Absorption & Neutral
\end{tabular}
\end{center}

\bigskip

The Wick-ordered Dyson series is then given by the sum 
\begin{equation}
U_{t}^{\left( \lambda \right) }=\sum_{\frak{D}}\frak{\hat{D}}_{t}\left(
\lambda \right)
\end{equation}
which we now describe. We sum over all diagrams $\frak{D}$ obtained by
writing $n$ vertices in a line $\left( n=0,1,2,\cdots \right) $ as below

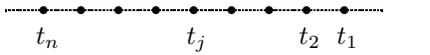
\begin{center}
%
%
%
%
%
%
\setlength{\unitlength}{.05cm}
\begin{picture}(120,20)
\label{pic1}

\put(20,10){\circle*{2}}
\put(30,10){\circle*{2}}
\put(40,10){\circle*{2}}
\put(50,10){\circle*{2}}
\put(60,10){\circle*{2}}
\put(70,10){\circle*{2}}
\put(80,10){\circle*{2}}
\put(90,10){\circle*{2}}
\put(100,10){\circle*{2}}

\put(18,1){$t_{n}$}
\put(88,1){$t_{2}$}
\put(98,1){$t_{1}$}
\put(58,1){$t_{j}$}

\put(10,10){\dashbox{0.5}(100,0){ }}

\end{picture}
%
,
\end{center}

\bigskip

\noindent taking each of the vertices to be one of the four shown above,
connecting up some of the creation/annihilation pairs and leaving the rest
as external lines.

With each such diagram $\frak{D}$ we associate the operator 
\begin{equation}
\frak{\hat{D}}_{t}\left( \lambda \right) =\left( -\mathrm{i}\right)
^{n}E_{n}\cdots E_{1}\otimes \int_{\Delta _{n}\left( t\right) }\Pi a^{\dag
}\left( \lambda \right) \Pi C\left( \lambda \right) \Pi a\left( \lambda
\right)
\end{equation}
where, depending on the type of vertex at $t_{j}$, 
\begin{equation*}
E_{j}=\left\{ 
\begin{array}{c}
E_{ij},\text{ scattering,} \\ 
E_{i0},\text{ emission,} \\ 
E_{0j},\text{ absorption,} \\ 
E_{00},\text{ neutral.}
\end{array}
\right.
\end{equation*}
and in the simplex integral $\Pi a^{\dag }\left( \lambda \right) $ is the
product over all external lines going out, $\Pi a\left( \lambda \right) $ is
the product over all external lines coming in and $\Pi C\left( \lambda
\right) $ is a product over all contraction pairs.

For instance the nine vertex diagram below

\bigskip

\bigskip

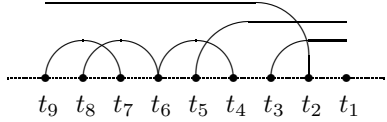
\begin{center}
%
%
%
%
%
%
%
%
\setlength{\unitlength}{.05cm}
\begin{picture}(120,20)
\label{pic1}

\put(20,10){\circle*{2}}
\put(30,10){\circle*{2}}
\put(40,10){\circle*{2}}
\put(50,10){\circle*{2}}
\put(60,10){\circle*{2}}
\put(70,10){\circle*{2}}
\put(80,10){\circle*{2}}
\put(90,10){\circle*{2}}
\put(100,10){\circle*{2}}

\put(30,10){\oval(20,20)[t]}
\put(40,10){\oval(20,20)[t]}
\put(60,10){\oval(20,20)[t]}
\put(20,10){\oval(140,40)[tr]}
\put(100,10){\oval(40,20)[tl]}
\put(100,10){\oval(80,30)[tl]}

\put(18,1){$t_{9}$}
\put(28,1){$t_{8}$}
\put(38,1){$t_{7}$}
\put(48,1){$t_{6}$}
\put(58,1){$t_{5}$}
\put(68,1){$t_{4}$}
\put(78,1){$t_{3}$}
\put(88,1){$t_{2}$}
\put(98,1){$t_{1}$}

\put(10,10){\dashbox{0.5}(100,0){ }}

\end{picture}
%
\end{center}

\noindent corresponds to the operator 
\begin{eqnarray*}
&&-\mathrm{i}%
E_{0j_{9}}E_{0j_{8}}E_{i_{7}0}E_{i_{6}j_{6}}E_{0j_{5}}E_{i_{4}0}E_{0j_{3}}E_{i_{2}0}E_{00}
\\
&&\otimes \int_{\Delta _{9}\left( t\right) }a_{i_{2}}^{\dag }\left(
t_{2},\lambda \right) a_{j_{5}}\left( t_{5},\lambda \right) a_{j_{3}}\left(
t_{3},\lambda \right) \\
&&C_{j_{9}i_{7}}\left( t_{9}-t_{7},\lambda \right) C_{j_{8}i_{6}}\left(
t_{8}-t_{6},\lambda \right) C_{j_{6}i_{4}}\left( t_{6}-t_{4},\lambda \right)
.
\end{eqnarray*}

The diagrams we are considering are Goldstone diagrams, or time-ordered
Feynman diagrams \cite{Mattuck}. The vertices however carry an operator
weight $E_{\alpha \beta }$ and, even in the absence of external lines, the
diagrams will not generally be scalars!

\textbf{Step 2: The Markov Limit of Individual Diagrams}

Let us consider the effect of the Markov limit $\lambda \rightarrow 0$ on
individual diagrams. If we have a contraction

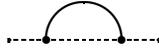
\begin{center}
%
%
%
%
\setlength{\unitlength}{.1cm}
\begin{picture}(20,5)
\label{picc}

\put(0,0){\dashbox{0.5}(20,0){ }}

\thicklines

\put(5,0){\circle*{1}}
\put(15,0){\circle*{1}}
\put(10,0){\oval(10,10)[t]}

\end{picture}
%
\end{center}

\noindent over consecutive times $t_{j+1}>t_{j}$, then we will have $%
t_{j+1}-t_{j}\rightarrow 0^{+}$ in the limit. The effect of each such
contraction is to reduce the order of the simplex by one degree and to
introduce a multiplication factor $V^{ij}$. (Remember, we only get a partial
contribution from the delta-function since $t_{j+1}>t_{j}$.) \ On the other
hand, if any of the contracted time pairs are not time consecutive then the
contribution converges weakly to zero! The reason is essentially that not
just the emission and absorption times, but all the intermediate times are
forced to be equal, and we get a collapse of the integral to zero.

We therefore say that a diagram is time-consecutive (TC) if each contraction
appearing is between time consecutive vertices only.

\textbf{Step 3: The Vacuum Limit}

We now fix $u,v\in \frak{h}$ and investigate the limit$\ \lim_{\lambda
\rightarrow 0}\left\langle u\otimes \Omega |U_{t}^{\left( \lambda \right)
}|v\otimes \Omega \right\rangle $ which we denote by the single vertex

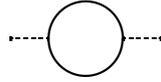
\begin{center}
%
%
%
%
%
\setlength{\unitlength}{.1cm}
\begin{picture}(20,5)
\label{picc}

\put(0,0){\dashbox{0.5}(5,0){ }}
\put(15,0){\dashbox{0.5}(5,0){ }}

\thicklines

\put(10,0){\circle{10}}

\end{picture}
%

\bigskip
\end{center}

\noindent We now argue that this limit will consist of the TC diagrams
having no external lines which we can express as

\begin{center}
%
%
%
%
%
\setlength{\unitlength}{.05cm}
\begin{picture}(20,5)
\label{picc}

\put(0,0){\dashbox{0.5}(5,0){ }}
\put(15,0){\dashbox{0.5}(5,0){ }}

\thicklines

\put(10,0){\circle{10}}

\end{picture}
%
=%
%
%
%
\setlength{\unitlength}{.05cm}
\begin{picture}(20,5)
\label{picc}

\put(0,0){\dashbox{0.5}(20,0){ }}

\end{picture}
%
+%
%
%
%
\setlength{\unitlength}{.05cm}
\begin{picture}(20,5)
\label{picc}

\put(0,0){\dashbox{0.5}(20,0){ }}

\thicklines
\put(10,0){\circle*{2}}

\end{picture}
%
+(%
%
%
%
\setlength{\unitlength}{.05cm}
\begin{picture}(20,5)
\label{picc}

\put(0,0){\dashbox{0.5}(20,0){ }}

\thicklines
\put(5,0){\circle*{2}}
\put(15,0){\circle*{2}}

\end{picture}
%
+%
%
%
%
%
\setlength{\unitlength}{.05cm}
\begin{picture}(20,5)
\label{picc}

\put(0,0){\dashbox{0.5}(20,0){ }}

\thicklines
\put(5,0){\circle*{2}}
\put(15,0){\circle*{2}}

\put(10,0){\oval(10,10)[t]}

\end{picture}
%
)

\bigskip

+(%
%
%
%
\setlength{\unitlength}{.05cm}
\begin{picture}(30,5)
\label{picc}

\put(0,0){\dashbox{0.5}(30,0){ }}

\thicklines
\put(5,0){\circle*{2}}
\put(15,0){\circle*{2}}
\put(25,0){\circle*{2}}

\end{picture}
%
+%
%
%
%
%
%
\setlength{\unitlength}{.05cm}
\begin{picture}(30,5)
\label{picc}

\put(0,0){\dashbox{0.5}(30,0){ }}

\thicklines
\put(5,0){\circle*{2}}
\put(15,0){\circle*{2}}
\put(25,0){\circle*{2}}

\put(20,0){\oval(10,10)[t]}

\end{picture}
%
+%
%
%
%
%
\setlength{\unitlength}{.05cm}
\begin{picture}(30,5)
\label{picc}

\put(0,0){\dashbox{0.5}(30,0){ }}

\thicklines
\put(5,0){\circle*{2}}
\put(15,0){\circle*{2}}
\put(25,0){\circle*{2}}

\put(10,0){\oval(10,10)[t]}

\end{picture}
%
+%
%
%
%
%
\setlength{\unitlength}{.05cm}
\begin{picture}(30,5)
\label{picc}

\put(0,0){\dashbox{0.5}(30,0){ }}

\thicklines
\put(5,0){\circle*{2}}
\put(15,0){\circle*{2}}
\put(25,0){\circle*{2}}

\put(10,0){\oval(10,10)[t]}
\put(20,0){\oval(10,10)[t]}

\end{picture}
%
)+ $\ \cdots $

\bigskip
\end{center}

The diagrams have been grouped by vertex number, however, it also possible
to group them by effective vertex number which is the actual number minus
the number of contractions and which gives reduced degree of the simplex.
The series can be partially re-summed as

\begin{center}
%
%
%
%
%
\setlength{\unitlength}{.05cm}
\begin{picture}(20,5)
\label{picc}

\put(0,0){\dashbox{0.5}(5,0){ }}
\put(15,0){\dashbox{0.5}(5,0){ }}

\thicklines

\put(10,0){\circle{10}}

\end{picture}
%
=%
%
%
%
\setlength{\unitlength}{.05cm}
\begin{picture}(20,5)
\label{picc}

\put(0,0){\dashbox{0.5}(20,0){ }}

\end{picture}
%
+%
%
%
%
%
\setlength{\unitlength}{.05cm}
\begin{picture}(20,5)
\label{picc}

\put(0,0){\dashbox{0.5}(5,0){ }}

\put(5,-5){\line(1,0){10}}
\put(15,-5){\line(0,1){10}}
\put(15,5){\line(-1,0){10}}
\put(5,5){\line(0,-1){10}}

\put(15,0){\dashbox{0.5}(5,0){ }}

\end{picture}
%
+%
%
%
%
%
%
%
\setlength{\unitlength}{.05cm}
\begin{picture}(40,5)
\label{picc}

\put(0,0){\dashbox{0.5}(5,0){ }}

\put(5,-5){\line(1,0){10}}
\put(15,-5){\line(0,1){10}}
\put(15,5){\line(-1,0){10}}
\put(5,5){\line(0,-1){10}}

\put(15,0){\dashbox{0.5}(5,0){ }}

\put(20,-5){\line(1,0){10}}
\put(30,-5){\line(0,1){10}}
\put(30,5){\line(-1,0){10}}
\put(20,5){\line(0,-1){10}}

\put(30,0){\dashbox{0.5}(5,0){ }}

\end{picture}
%
+%
%
%
%
%
%
%
%
%
%
%
\setlength{\unitlength}{.05cm}
\begin{picture}(50,5)
\label{picc}

\put(0,0){\dashbox{0.5}(5,0){ }}

\put(5,-5){\line(1,0){10}}
\put(15,-5){\line(0,1){10}}
\put(15,5){\line(-1,0){10}}
\put(5,5){\line(0,-1){10}}

\put(15,0){\dashbox{0.5}(5,0){ }}

\put(20,-5){\line(1,0){10}}
\put(30,-5){\line(0,1){10}}
\put(30,5){\line(-1,0){10}}
\put(20,5){\line(0,-1){10}}

\put(30,0){\dashbox{0.5}(5,0){ }}

\put(35,-5){\line(1,0){10}}
\put(45,-5){\line(0,1){10}}
\put(45,5){\line(-1,0){10}}
\put(35,5){\line(0,-1){10}}

\put(45,0){\dashbox{0.5}(5,0){ }}

\end{picture}
%
+ $\ \ \cdots $
\end{center}

\noindent where each box is a sum over all effective one-vertex
contributions:

\bigskip

\begin{center}
%
%
%
%
%
%
\setlength{\unitlength}{.05cm}
\begin{picture}(20,5)
\label{picc}

\put(5,-5){\line(1,0){10}}
\put(15,-5){\line(0,1){10}}
\put(15,5){\line(-1,0){10}}
\put(5,5){\line(0,-1){10}}

\end{picture}
%
=%
%
%
%
%
\setlength{\unitlength}{.05cm}
\begin{picture}(10,5)
\label{picc}

\thicklines
\put(5,0){\circle*{2}}

\end{picture}
%
+%
%
%
%
%
\setlength{\unitlength}{.05cm}
\begin{picture}(10,5)
\label{picc}

\put(0,0){\dashbox{0.5}(10,0){ }}

\thicklines
\put(0,0){\circle*{2}}
\put(10,0){\circle*{2}}

\put(5,0){\oval(10,10)[t]}

\end{picture}
%
+%
%
%
%
%
\setlength{\unitlength}{.05cm}
\begin{picture}(20,5)
\label{picc}

\put(0,0){\dashbox{0.5}(20,0){ }}

\thicklines
\put(0,0){\circle*{2}}
\put(10,0){\circle*{2}}
\put(20,0){\circle*{2}}

\put(5,0){\oval(10,10)[t]}
\put(15,0){\oval(10,10)[t]}

\end{picture}
%
+%
%
%
%
%
\setlength{\unitlength}{.05cm}
\begin{picture}(30,5)
\label{picc}

\put(0,0){\dashbox{0.5}(30,0){ }}

\thicklines
\put(0,0){\circle*{2}}
\put(10,0){\circle*{2}}
\put(20,0){\circle*{2}}
\put(30,0){\circle*{2}}

\put(5,0){\oval(10,10)[t]}
\put(15,0){\oval(10,10)[t]}
\put(25,0){\oval(10,10)[t]}

\end{picture}
%
+%
%
%
%
%
\setlength{\unitlength}{.05cm}
\begin{picture}(40,5)
\label{picc}

\put(0,0){\dashbox{0.5}(40,0){ }}

\thicklines
\put(0,0){\circle*{2}}
\put(10,0){\circle*{2}}
\put(20,0){\circle*{2}}
\put(30,0){\circle*{2}}
\put(40,0){\circle*{2}}

\put(5,0){\oval(10,10)[t]}
\put(15,0){\oval(10,10)[t]}
\put(25,0){\oval(10,10)[t]}
\put(35,0){\oval(10,10)[t]}

\end{picture}
%
+ $\cdots $
\end{center}

\noindent which is analogous to the expression of the self-energy in quantum
field theory:as a sum over irreducible terms. (Note that series terminates
at second order when there is no scattering: as this is a form of cumulant
expansion, the emission/absorption problem is then Gaussian while allowing
scattering means that we have cumulant moments to all orders!) Explicitly,
the box at time vertex $t$ corresponds to the sum 
\begin{gather*}
-\mathrm{i}E_{00}dt+\left( -\mathrm{i}\right)
^{2}E_{0i_{1}}E_{i_{1}0}dt+\left( -\mathrm{i}\right)
^{3}E_{0i_{4}}V^{i_{4}i_{3}}E_{i_{3}i_{2}}V^{i_{2}i_{1}}E_{i_{1}0}dt+\cdots
\\
=\left( -\mathrm{i}E_{00}-E_{0i}V^{ij}\left( \frac{1}{1+\mathrm{i}\mathsf{VE}%
}\right) _{jk}E_{k0}\right) dt\equiv -\mathrm{i}G_{00}\,dt
\end{gather*}
and we have summed the geometric series, convergent since $\left\| \mathsf{VE%
}\right\| <1$, to get the required coefficient $G_{00}$. We can write the
expansion in the recursive form

\bigskip

\begin{center}
%
%
%
%
%
\setlength{\unitlength}{.05cm}
\begin{picture}(20,5)
\label{picc}

\put(0,0){\dashbox{0.5}(5,0){ }}
\put(15,0){\dashbox{0.5}(5,0){ }}

\thicklines

\put(10,0){\circle{10}}

\end{picture}
%
=%
%
%
%
\setlength{\unitlength}{.05cm}
\begin{picture}(20,5)
\label{picc}

\put(0,0){\dashbox{0.5}(20,0){ }}

\end{picture}
%
+%
%
%
%
%
%
%
\setlength{\unitlength}{.05cm}
\begin{picture}(40,5)
\label{picc}

\put(0,0){\dashbox{0.5}(5,0){ }}

\put(5,-5){\line(1,0){10}}
\put(15,-5){\line(0,1){10}}
\put(15,5){\line(-1,0){10}}
\put(5,5){\line(0,-1){10}}

\put(15,0){\dashbox{0.5}(5,0){ }}

\put(30,0){\dashbox{0.5}(5,0){ }}

\thicklines
\put(25,0){\circle{10}}

\end{picture}
%

\bigskip
\end{center}

\noindent and this is interpreted as the integro-differential equation 
\begin{equation*}
\int_{0}^{t}\left\langle u\otimes \Omega |dU_{t_{1}}|v\otimes \Omega
\right\rangle =\left\langle u\otimes \Omega |v\otimes \Omega \right\rangle -%
\mathrm{i}\int_{0}^{t}dt_{2}\left\langle u\otimes \Omega
|G_{00}\int_{0}^{t_{2}}dU_{t_{1}}|v\otimes \Omega \right\rangle
\end{equation*}
with decaying exponential solution 
\begin{equation*}
\left\langle u\otimes \Omega |U\left( t\right) |v\otimes \Omega
\right\rangle =\left\langle u|e^{-\mathrm{i}tG_{00}}|v\right\rangle .
\end{equation*}

The interpretation is intended to suggest that there is a limit object $%
U_{t} $ which we interpret as a unitary quantum stochastic process on a
noise space with initial space $\frak{h}$.

\textbf{Step 4: Limit of Exponential Vector Matrix Elements}

Calculating $\lim_{\lambda \rightarrow 0}\left\langle u\otimes \varepsilon
_{\lambda }\left( \mathbf{f}\right) |U\left( t,\lambda \right) |v\otimes
\varepsilon _{\lambda }\left( \mathbf{g}\right) \right\rangle $ does not
require too much machinery beyond that used in the vacuum case. Indeed, we
can convert the new matrix elements to vacuum ones using the relation 
\begin{equation*}
\left\langle u\otimes \varepsilon _{\lambda }\left( \mathbf{f}\right)
|U_{t}^{\left( \lambda \right) }|v\otimes \varepsilon _{\lambda }\left( 
\mathbf{g}\right) \right\rangle =\left\langle u\otimes \Omega |\tilde{U}%
_{t}^{\left( \lambda \right) }|v\otimes \Omega \right\rangle
\end{equation*}
where $\tilde{U}_{t}^{\left( \lambda \right) }$ is the solution to the
Schr\"{o}dinger equation determined by the ``Hamiltonian'' $\tilde{\Upsilon}%
_{t}^{\left( \lambda \right) }$ obtained by replacing the fields $a\left(
t,\lambda \right) $ and $a^{\dag }\left( t,\lambda \right) $ with 
\begin{eqnarray*}
\tilde{a}_{i}\left( t,\lambda \right) &=&a_{i}\left( t,\lambda \right) +\int
C_{ij}\left( t-s,\lambda \right) g_{j}\left( s\right) , \\
\tilde{a}_{i}^{\dagger }\left( t,\lambda \right) &=&a_{i}^{\dagger }\left(
t,\lambda \right) +\left[ \int C_{ij}\left( t-s\right) f_{j}\left( t\right) %
\right] ^{\ast }.
\end{eqnarray*}
The new interaction $\tilde{\Upsilon}_{t}^{\left( \lambda \right) }$ isn't
necessarily self-adjoint, however, that doesn't effect things. We may
rearrange $\tilde{\Upsilon}_{t}^{\left( \lambda \right) }$ in terms of the
original fields as $\tilde{E}_{\alpha \beta }\otimes a_{\alpha }^{\dag
}\left( t,\lambda \right) a_{\beta }\left( t,\lambda \right) $ where,
suppressing the $\lambda $ and $t$ dependences, the $\tilde{E}_{\alpha \beta
}$ are the operators 
\begin{eqnarray*}
\tilde{E}_{ij} &=&E_{ij}, \\
\tilde{E}_{i0} &=&E_{i0}+E_{ij}\int C_{jl}\left( t-s,\lambda \right)
g_{l}\left( s\right) , \\
\tilde{E}_{0j} &=&E_{0j}+E_{jk}\left[ \int C_{kl}\left( t-s\right)
f_{l}\left( t\right) \right] ^{\ast }, \\
\tilde{E}_{00} &=&E_{00}+E_{i0}\left[ \int C_{il}\left( t-s\right)
f_{l}\left( t\right) \right] ^{\ast }+E_{0j}\int C_{jl}\left( t-s,\lambda
\right) g_{l}\left( s\right) \\
&&+E_{jk}\left[ \int C_{il}\left( t-s\right) f_{l}\left( t\right) \right]
^{\ast }\int C_{jm}\left( t-s,\lambda \right) g_{m}\left( s\right) .
\end{eqnarray*}

Therefore we only have to repeat our previous argument, but with the
original coefficients now replaced by the modified ones $\tilde{E}_{\alpha
\beta }$, taking care with the $t$ and $\lambda $ dependences. This time the
box at time vertex $t$ corresponds to the sum $-\mathrm{i}f_{\alpha }^{\ast
}g_{\beta }G_{\alpha \beta }dt$ where now 
\begin{eqnarray*}
G_{\alpha \beta } &=&E_{\alpha \beta }-E_{\alpha i}\left( \mathsf{V}-\mathrm{%
i}\mathsf{VEV}+\left( -\mathrm{i}\right) ^{2}\mathsf{VEVEV}+\cdots \right)
_{ij}E_{j\beta } \\
&=&\left( -\mathrm{i}E_{\alpha \beta }-E_{\alpha i}V^{ij}\left( \frac{1}{1+%
\mathrm{i}\mathsf{VE}}\right) _{jk}E_{k\beta }\right)
\end{eqnarray*}
and this gives the required result.

\textbf{Step 5: Convergence of the Series}

What we have done\ so far has been to expand the Dyson series, determine the
asymptotic limit of each diagram term (only the TC ones survived), to
replace the terms by their respective limits and to re-sum the series. To
complete the proof, we need to establish that the series is absolutely and
uniformly convergent. Fortunately we are able to extend proof for estimating
these type of series exists for emission-absorption interactions \cite{Pule}
to the general case.

\bigskip

Let us start with the case where we have emission and absorption only in the
interaction. The order must be even say, $n=2n_{2}$, as the vacuum diagrams
consist of $n_{2}$ pair contractions only. There will then be $\frac{\left(
2n_{2}\right) !}{2^{n_{2}}n_{2}!}$ such diagrams with $2n_{2}$ vertices. A
typical diagram, having $n_{2}=6$ is sketched below:

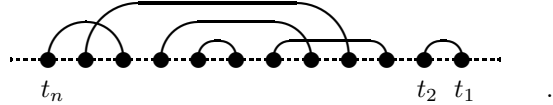
\begin{center}
%
%
%
%
\setlength{\unitlength}{.1cm}
\begin{picture}(70,20)
\label{picG4}

\put(0,5){\dashbox{0.5}(65,0){ }}

\thicklines

\put(5,5){\circle*{2}}
\put(10,5){\circle*{2}}
\put(15,5){\circle*{2}}
\put(20,5){\circle*{2}}
\put(25,5){\circle*{2}}
\put(30,5){\circle*{2}}
\put(35,5){\circle*{2}}
\put(40,5){\circle*{2}}
\put(45,5){\circle*{2}}
\put(50,5){\circle*{2}}
\put(55,5){\circle*{2}}
\put(60,5){\circle*{2}}

\put(10,5){\oval(10,10)[t]}
\put(57.5,5){\oval(5,5)[t]}
\put(42.5,5){\oval(15,5)[t]}
\put(27.5,5){\oval(5,5)[t]}
\put(30,5){\oval(20,10)[t]}
\put(27.5,5){\oval(35,15)[t]}
\put(59,0){$t_1$}
\put(54,0){$t_2$}
\put(4,0){$t_n$}
\end{picture}
%
.
\end{center}

There exists a permutation $\sigma $ of the $n=2n_{2}$ time indices which
re-orders to the\ diagram $\frak{D}_{0}\left( n\right) $\ shown below

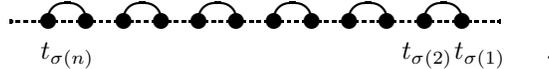
\begin{center}
%
%
%
%
%
\setlength{\unitlength}{.1cm}
\begin{picture}(70,10)
\label{picG5}

\put(0,5){\dashbox{0.5}(65,0){ }}

\thicklines

\put(5,5){\circle*{2}}
\put(10,5){\circle*{2}}
\put(15,5){\circle*{2}}
\put(20,5){\circle*{2}}
\put(25,5){\circle*{2}}
\put(30,5){\circle*{2}}
\put(35,5){\circle*{2}}
\put(40,5){\circle*{2}}
\put(45,5){\circle*{2}}
\put(50,5){\circle*{2}}
\put(55,5){\circle*{2}}
\put(60,5){\circle*{2}}

\put(47.5,5){\oval(5,5)[t]}
\put(57.5,5){\oval(5,5)[t]}
\put(37.5,5){\oval(5,5)[t]}
\put(27.5,5){\oval(5,5)[t]}
\put(17.5,5){\oval(5,5)[t]}
\put(7.5,5){\oval(5,5)[t]}

\put(59,0){$t_{\sigma (1)}$}
\put(52,0){$t_{\sigma (2)}$}
\put(4,0){$t_{\sigma (n)}$}
\end{picture}
%
.
\end{center}

The permutation is moreover unique if it has the induced ordering of the
emission times. Not all permutations arise this way, the ones that do are
termed admissible. We now consider an estimate of the $n-$th term in the
Dyson series. Let $E=\max \left\| E_{\alpha \beta }\right\| $, then 
\begin{eqnarray*}
\sum_{\frak{D}}E^{n}\int_{\Delta _{n}\left( t\right) }\left. \prod \left|
C\left( \lambda \right) \right| \right| _{\frak{D}} &=&\sum_{\text{%
Admissible perms.}}E^{n}\int_{\Delta _{n}\left( t\right) }\left. \prod
\left| C\left( \lambda \right) \circ \sigma \right| \right| _{\frak{D}%
_{0}\left( n\right) } \\
&=&E^{n}\int_{R\left( t\right) }\prod_{k=1}^{n_{2}}\left| C\left(
t_{2k}-t_{2k-1},\lambda \right) \right|
\end{eqnarray*}
where $R\left( t\right) $ is the union of simplices $\left\{ \left(
t_{n},\cdots ,t_{1}\right) :t>t_{\sigma ^{-1}\left( n\right) }>\cdots
>t_{\sigma ^{-1}\left( 1\right) }>0\right\} $ over all admissible
permutations $\mathsf{Z}$. $R\left( t\right) $ will be a subset of $\left[
0,t\right] ^{2n_{2}}$ and if we introduce variables $t_{2k}$ and $%
s_{2k}=t_{2k}-t_{2k-1}$ for $k=1,\cdots ,n_{2}$, it is easily seen that the
above is majorized by $E^{2n_{2}}C^{n_{2}}\times \dfrac{\max \left(
t,1\right) ^{n_{2}}}{n_{2}!}$, where $C=\max V^{ij}$. This is the Pul\`{e}
inequality \cite{Pule} and\ clearly gives the uniform absolute estimate
required to sum the series.

\bigskip

We now consider scattering, and constant, terms in the interaction. This
time, the number of diagrams with $n$ vertices will be given by the $n$-th
Bell number $B_{n}$. To see why this is so, we recall that if we have a
reservoir quanta created at a vertex, then perhaps scattered ,and finally
reabsorbed then we can think of it as the same quantum and treat all the
vertices it has been at as being linked. Each such diagram is then described
by these subsets of linked vertices (we should also count the neutral
vertices as these are singleton sets): in this way we have a one-to-one
correspondence between the diagrams and partitions of vertices into
non-empty subsets. The Bell numbers grow rapidly and have a complicated
asymptotic behavior. The proliferation of diagrams is due mainly to the
multiple scattering that now takes place.

Let us consider a typical diagram. We shall assume that within the diagram
there are $n_{1}$ singleton vertices [$\cdots $%
%
%
%
%
%
\setlength{\unitlength}{.1cm}
\begin{picture}(10,5)
\label{picea}

\put(0,0){\dashbox{0.5}(10,0){ }}

\thicklines

\put(5,0){\circle*{1}}

\end{picture}
%
$\cdots ]$, $n_{2}$ contraction pairs [$\cdots $%
%
%
%
\setlength{\unitlength}{.05cm}
\begin{picture}(26,5)
\label{picfa}

\put(0,0){\dashbox{0.5}(7,0){ }}
\put(19,0){\dashbox{0.5}(7,0){ }}
\put(9,0){$\cdots$}
\thicklines

\put(5,0){\circle*{2}}
\put(21,0){\circle*{2}}
\put(13,0){\oval(16,16)[t]}

\end{picture}
%
$\cdots ]$, $n_{3}$ contraction triples [$\cdots $%
%
%
%
%
%
\setlength{\unitlength}{.05cm}
\begin{picture}(42,8)
\label{picga}

\put(0,0){\dashbox{0.5}(7,0){ }}
\put(19,0){\dashbox{0.5}(4,0){ }}
\put(35,0){\dashbox{0.5}(7,0){ }}
\put(9,0){$\cdots$}
\put(25,0){$\cdots$}
\thicklines

\put(5,0){\circle*{2}}
\put(21,0){\circle*{2}}
\put(37,0){\circle*{2}}

\put(13,0){\oval(16,16)[t]}
\put(29,0){\oval(16,16)[t]}

\end{picture}
%
$\cdots ]$, etc. That is the diagram has a total of $n=\sum_{j}jn_{j}$
vertices which are partitioned into $m=\sum_{j}n_{j}$ connected subdiagrams.
For instance, we might have an initial segment of a diagram looking like the
following:

\bigskip

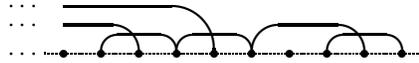
\begin{center}
%
%
%
%
%
%
%
%
%
%
\setlength{\unitlength}{.05cm}
\begin{picture}(110,20)
\label{picga}

\put(5,0){\dashbox{0.5}(100,0){ }}

\thicklines

\put(-4,0){\circle*{1}}
\put(-1,0){\circle*{1}}
\put(2,0){\circle*{1}}

\put(-4,7.5){\circle*{1}}
\put(-1,7.5){\circle*{1}}
\put(2,7.5){\circle*{1}}

\put(-4,12.5){\circle*{1}}
\put(-1,12.5){\circle*{1}}
\put(2,12.5){\circle*{1}}

\put(10,0){\circle*{2}}
\put(20,0){\circle*{2}}
\put(30,0){\circle*{2}}
\put(40,0){\circle*{2}}
\put(50,0){\circle*{2}}
\put(60,0){\circle*{2}}
\put(70,0){\circle*{2}}
\put(80,0){\circle*{2}}
\put(90,0){\circle*{2}}
\put(100,0){\circle*{2}}

\put(90,0){\oval(20,10)[t]}
\put(75,0){\oval(30,15)[t]}
\put(50,0){\oval(20,10)[t]}
\put(30,0){\oval(20,10)[t]}
\put(10,0){\oval(40,15)[tr]}
\put(10,0){\oval(80,25)[tr]}

\end{picture}
%
\end{center}

There will exist a permutation $\sigma $ of the $n$ vertices which will
reorder the vertices so that we have the singletons first, then the pair
contractions, then the triples, etc., so that we obtain a picture of the
following type

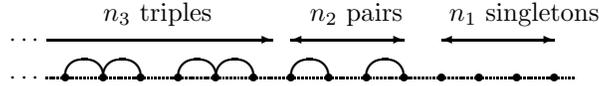
\begin{center}
%
%
%
%
%
%
%
%
%
%
%
%
%
%
\setlength{\unitlength}{.05cm}
\begin{picture}(150,20)
\label{picga}

\put(5,0){\dashbox{0.5}(140,0){ }}

\put(110,10){\vector(1,0){30}}
\put(140,10){\vector(-1,0){30}}
\put(112,15){$n_1$ singletons}

\put(70,10){\vector(1,0){30}}
\put(100,10){\vector(-1,0){30}}
\put(75,15){$n_2$ pairs}

\put(5,10){\vector(1,0){60}}
\put(20,15){$n_3$ triples}

\thicklines

\put(-4,0){\circle*{1}}
\put(-1,0){\circle*{1}}
\put(2,0){\circle*{1}}

\put(-4,10){\circle*{1}}
\put(-1,10){\circle*{1}}
\put(2,10){\circle*{1}}

\put(10,0){\circle*{2}}
\put(20,0){\circle*{2}}
\put(30,0){\circle*{2}}
\put(40,0){\circle*{2}}
\put(50,0){\circle*{2}}
\put(60,0){\circle*{2}}
\put(70,0){\circle*{2}}
\put(80,0){\circle*{2}}
\put(90,0){\circle*{2}}
\put(100,0){\circle*{2}}
\put(110,0){\circle*{2}}
\put(120,0){\circle*{2}}
\put(130,0){\circle*{2}}
\put(140,0){\circle*{2}}

\put(15,0){\oval(10,10)[t]}
\put(25,0){\oval(10,10)[t]}
\put(45,0){\oval(10,10)[t]}
\put(55,0){\oval(10,10)[t]}
\put(75,0){\oval(10,10)[t]}
\put(95,0){\oval(10,10)[t]}

\end{picture}
%
.
\end{center}

The permutation is again unique if we retain the induced ordering of the
first emission times for each connected block. We now wish to find a uniform
estimate for the $n$-th term in the Dyson series, we have 
\begin{equation*}
\sum_{\frak{D}}\int_{\Delta _{n}\left( t\right) }\prod \left| C\left(
\lambda \right) \right| \times \text{``weights''}
\end{equation*}
where the weights are the operator norms of various products of the type $%
E_{\alpha _{n}\beta _{n}}\cdots E_{\alpha _{1}\beta _{1}}$. Each connected
diagram of $j\geq 2$ vertices will typically have one emission and one
absorption, and $j-2$ scattering vertices. The Pul\`{e} argument of
rearranging the sum over diagrams into a single integral over a region $%
R\left( t\right) $ of $\left[ 0,t\right] ^{n}$ again applies and by similar
reason we arrive at the upper bound, this time of the type 
\begin{equation*}
\sum\nolimits_{n_{1},n_{2},n_{3},\cdots }^{\prime }\left\| \mathsf{VE}%
\right\| ^{n_{1}+2n_{2}+3n_{3}+\cdots }E^{n_{1}+n_{2}+n_{3}+\cdots }\times 
\frac{\max \left( t,1\right) ^{n_{1}+n_{2}+n_{3}+\cdots }}{%
n_{1}!n_{2}!n_{3}!\cdots }.
\end{equation*}
Here the sum is restricted so that $\sum_{j}jn_{j}=n$. An uniform estimate
for the entire series is then given by removing this restriction: 
\begin{equation*}
\Xi \left( A,B\right) =\sum_{n_{1},n_{2},n_{3},\cdots }\frac{\exp \left\{
\sum_{j}\left( Aj+B\right) n_{j}\right\} }{n_{1}!n_{2}!n_{3}!\cdots }
\end{equation*}
where $e^{A}=\left\| \mathsf{VE}\right\| $ and $e^{B}=E\max \left(
t,1\right) $. Again we use the trick to convert a sum of products into a
product of sums 
\begin{equation*}
\Xi \left( A,B\right) =\prod_{j}\sum_{n}\frac{\exp \left\{ \left(
Aj+B\right) n\right\} }{n!}=\exp \left\{ \frac{e^{A+B}}{1-e^{A}}\right\} ,
\end{equation*}
where we need $e^{A}<1$ to sum the geometric series - this however, is
precisely our condition that $\left\| \mathsf{VE}\right\| <1$.

\textbf{Acknowledgment}

It is a great pleasure for the author to thank the staff at MabuchiLab for
their kind hospitality during his visit there when part of this paper was
written. Conversations with a Hideo Mabuchi, Gerard Milburn, P.
Krishnaprasad, Matthew James, Howard Wiseman, Andrew Doherty, Ramon van
Handel and Luc Bouten are gratefully acknowledged.

\end{document}